\documentclass[final,3p,times]{elsarticle}

\usepackage{graphicx}
\usepackage{multirow}
\usepackage{caption}
\usepackage{subcaption}
\usepackage{xcolor}
\usepackage{algorithm}
\usepackage{algorithmicx}
\usepackage{algpseudocode}
\usepackage{amsmath,amssymb,amsfonts,mathtools}
\usepackage{amsthm}
\usepackage{physics}
\usepackage{siunitx}
\usepackage{ulem}
\usepackage{tikz}
\usetikzlibrary{matrix}
\newcommand{\bb}[1]{\mathbf{#1}}
\newcommand{\bs}[1]{\boldsymbol{#1}}
\newcommand{\C}{\mathbb{C}}
\newcommand{\Comsol}{COMSOL Multiphysics\textsuperscript{\textregistered}}
\newcommand{\Matlab}{Matlab\textsuperscript{\textregistered}}

\usepackage{natbib}
\usepackage{hyperref}

\usepackage{todonotes}
\definecolor{ReviewGreen}{RGB}{0,180,0}

\journal{Journal of Sound and Vibration}

\begin{document}

\begin{frontmatter}



\title{Topology optimization of pentamode metamaterials for underwater acoustics}

\author[polimi,LMA]{Sebastiano Cominelli\corref{cor}} \cortext[cor]{sebastiano.cominelli@univ-amu.fr}
\author[polimi]{Matteo Pozzi} 
\author[polimi]{Francesco Braghin} 

\affiliation[polimi]{
    organization={Department of Mechanical Engineering, Politecnico di Milano},
    addressline={Via G.\ La Masa, 1}, 
    city={Milan},
    postcode={20156}, 
    country={Italy},
}
\affiliation[LMA]{
    organization={Aix-Marseille Université, Laboratoire de Méchanique et d'Acoustique},
    addressline={4 impasse Nikola Tesla}, 
    city={Marseille},
    postcode={13013}, 
    country={France},
}

\begin{abstract}
    This study presents an automated topology optimization framework for designing pentamode acoustic metamaterials. It provides precise control over the material effective acoustic properties while minimizing the shear modulus to achieve fluid-like behavior. The approach combines low-frequency homogenization for accurate property evaluation and the adjoint method for efficient sensitivity analysis. The Virtual Temperature Method (VTM) ensures structural connectivity and manufacturability, addressing the typical challenges of low-stiffness, high-mass-density microstructures. The framework is demonstrated through the design of a L\"uneburg lens and an acoustic invisibility cloak for underwater applications. Acoustic-elastic simulations validate the performance of both the unit cells and the complete devices. This method eliminates the need for predefined geometries, offering a flexible, reliable, and scalable alternative to conventional parametric optimization. It provides a powerful tool for the automated design of complex, anisotropic, pentamode metamaterials.
\end{abstract}

\begin{graphicalabstract}
    \includegraphics[width=\linewidth]{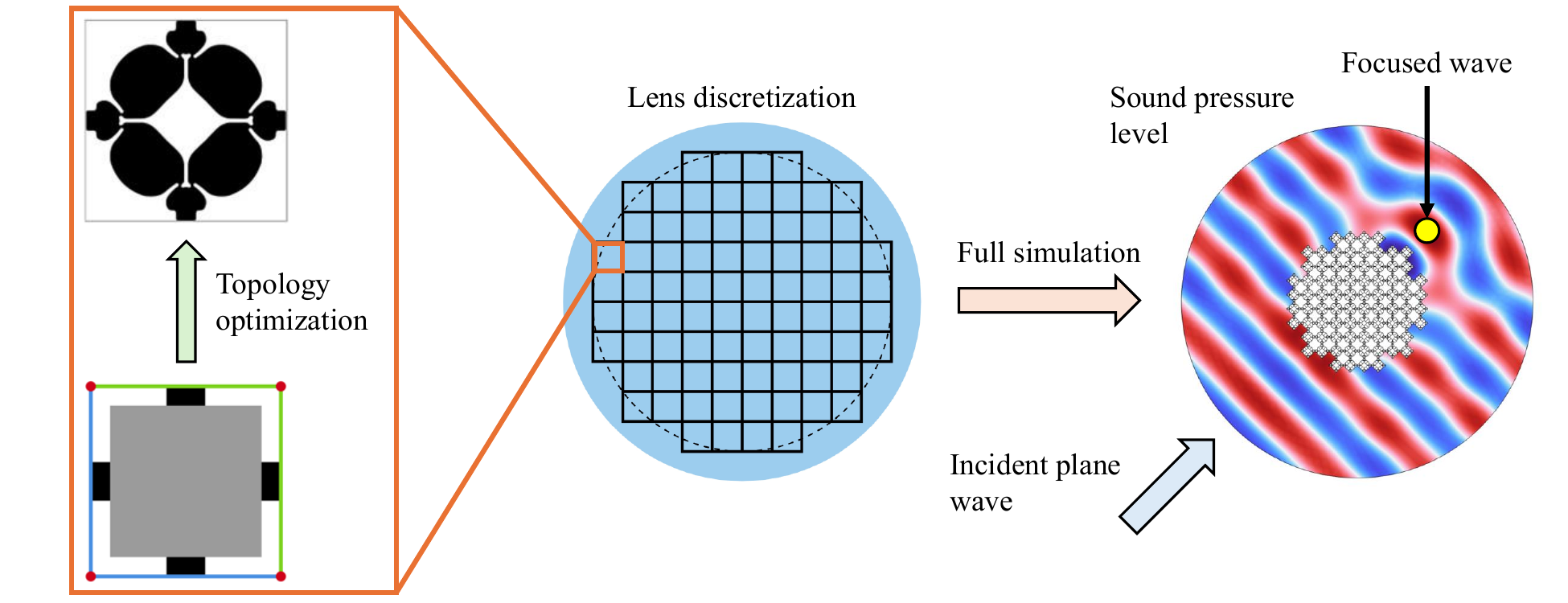}
\end{graphicalabstract}

\begin{highlights}
    \item Automated topology optimization framework for pentamode metamaterial design;
    \item Virtual Temperature Method to control thin connections and structural integrity;
    \item Optimized and validated designs for a Lüneburg lens and an underwater acoustic cloak.
\end{highlights}

\begin{keyword}
    Pentamode metamaterials \sep Topology optimization \sep Low-frequency homogenization \sep Underwater acoustics \sep Acoustic cloaking
\end{keyword}

\end{frontmatter}



\section{Introduction} \label{sec:introduction}


Acoustic metamaterials have revolutionized our ability to shape and guide sound, offering possibilities far beyond those of traditional materials. This progress is especially significant underwater, where long-distance communication almost entirely relies on acoustics. Over the past two decades, the development of acoustic metamaterials has led to many novel devices for wave manipulation, including acoustic lenses \cite{ lin2009gradient, climente2010sound, martin2010sonic, hladky2013negative, li2014three, ma2022underwater, allam20203d, li2025underwater}, invisibility cloaks \cite{chen2017broadband, bi2018experimental, Quadrelli2021, cominelli2022design}, and wave guides \cite{chen2022broadband}.
These devices are often made using materials with smoothly varying properties, which are obtained by assembling a microstructure whose geometry varies across space and is engineered to produce the desired macroscopic properties.
Several methods exist to design such a microstructure, but the most widespread is parametric optimization. This approach begins by selecting a base geometry that already exhibits some of the desired physical features. The geometry is then parameterized so that it can be adapted across the device, ideally spanning the full range of required properties. Finally, an optimization procedure tunes these parameters cell by cell (see, for example, \cite{chen2017broadband,Quadrelli2021,Luo2022pentamode,cominelli2022design,cominelli2023optimal,allam20203d}).
\\
Although effective, this workflow has inherent limitations. First, selecting a suitable base geometry is often challenging, and a single structure may not cover the full property space needed for the target device. Additionally, certain combinations of parameters may produce geometric conflicts or configurations that are physically impossible, making the process delicate and computationally expensive. Ultimately, this approach limits the attainable properties and the performance of the final device.

Topology optimization offers an alternative design route in which the mapping from the required effective properties to the underlying geometry can be, in principle, fully automated \cite{Bendsoe1988homogenization, Bendsoe2004}. Unlike parametric approaches, it removes the need for predefined base geometries and avoids conflicts arising from incompatible design parameters.
\\
This technique has been widely adopted to design periodic structures with tailored elastic \cite{Andreassen2014a, Andreassen2014b, Esfarjani2022topology}, thermal \cite{Sha2024topology}, and fluidic \cite{Osanov2016topology} responses. It enabled the creation of elastic metamaterials with unconventional or extremal behaviors, including auxetic lattices \cite{Gao2019auxetic}, phononic and photonic crystals \cite{Li2019phononic, Swartz2021photonic, Swartz2022periodic}, and pentamode architectures \cite{Wang2022topological, Zhang2020pentamode, W2021pentamode}.
Among the various optimization frameworks \cite{Xie1993eso, Allaire2004levelset, Wang2003, Guo2014mmc}, the density-based method \cite{Bendsoe1988homogenization} combined with the Solid Isotropic Material with Penalization (SIMP) scheme \cite{Bendsoe1999} allows flexibility in managing symmetries within the design domain. This strategy has played a central role in metamaterial design since the work of \citet{Sigmund1994homogenization, Sigmund1995tailoring, Sigmund2000extremal}, and is usually combined with homogenization theory \cite{Hassani1998a, Hassani1998b, Hassani1998c} to engineer periodic unit cells with prescribed effective properties.

Within the broad class of acoustic metamaterials, pentamode materials (PMs) have emerged as a particularly versatile platform. First identified in the search for extreme elastic properties \cite{milton1995which,Sigmund1995tailoring,Sigmund2000extremal}, PMs are elastic solids with five out of six easy modes of deformation. When their bulk modulus is much higher than their shear modulus, they exhibit acoustic behavior similar to fluids. These features make PMs well-suited for underwater acoustics, where they enable the design of artificial fluids with prescribed effective properties \cite{norris2008acoustic,norris2011metal,Kadic2012,brambilla2025high}. PMs offer advantages that are often difficult to achieve with most phononic crystals, such as broadband functionality, tunable anisotropy, and structural self-support in two and three dimensions.
\\
Designing PMs, however, remains particularly challenging. Their extreme contrast between bulk and shear moduli is usually achieved through thin elements that approximate the kinematics of hinges and sliders. Thus, a compromise with manufacturability is necessary.
Additionally, a precise filling fraction is usually required due to acoustic requirements on the average density. Therefore, the mass must be distributed within the cell without interfering with the functional components \cite{Kadic2012,brambilla2025high}. 
This combination of requirements is known to produce thin or disconnected members when using topology optimization algorithms \cite{Cool2025connectivity, Bonaccorsi2025connectivity}, and must be carefully monitored.
\\
Only a limited number of studies have attempted to optimize the topology of PMs, and these have largely focused on maximizing the ratio between bulk and shear moduli \cite{Luo2022pentamode}. Such approaches do not typically enforce specific target values for these moduli, despite the fact that precise elastic properties are required for acoustics. The works that address this need rely on a small number of design variables \cite{dong2021customized, rodriguez2024automated}, concentrate primarily on isotropic PMs \cite{Zhang2020pentamode}, or use topology optimization to generate an initial guess that is then refined through parametric optimization \cite{Luo2022pentamode, W2021pentamode}. With this work, we present a reliable, versatile, automated, and scalable procedure that is missing from the literature.
\\
In a preliminary study \cite{Pozzi2024luneburg}, we designed a PM-based L\"uneburg acoustic lens through topology optimization, using the Virtual Temperature Method (VTM) \cite{Liu2015, Li2016} to promote structural connectivity during optimization.
Building upon those promising initial results, the present work significantly refines the VTM implementation, simplifying the resulting microstructure. More importantly, we generalize the framework to enable the design of anisotropic graded-index PMs. We demonstrate this capability by developing an acoustic invisibility cloak.

The rest of the paper is organized as follows. In Section~\ref{sec:pentamode}, we introduces the elastic properties of PMs, outlining their significance in controlling acoustic waves. Then, a brief overview of low-frequency homogenization is presented to introduce the notation and the approach used to compute the effective properties of a unit cell. Section~\ref{sec:topopt} presents the topology optimization framework and details the strategy employed to avoid connectivity issues in the optimized layouts. This methodology is applied to designing a L\"uneburg lens in Section~\ref{sec:lens}, and an acoustic cloak in Section~\ref{sec:cloak}. Both case studies are validated through the numerical simulation of the unit cells, the effective properties and the microstructured devices. Finally, Section~\ref{sec:conclusion} draws conclusions and discusses potential extensions of the proposed framework.

 

\section{A pentamode lens} \label{sec:pentamode}

The L\"uneburg lens is a classic example of a gradient-index (GRIN) device. Its spherical refractive index profile is designed to focus incoming plane waves onto a point on the opposite side of its surface without aberration.  Originally developed in the context of optics \cite{luneburg1966mathematical}, the design of the L\"uneburg lens has been continuously refined for use in underwater acoustics (see, for example, \cite{allam20203d,tong20233d,lee2025underwater,li2025underwater}).
In the following, we briefly summarize the conceptual steps we propose to construct such a lens. This example illustrates a more general procedure for designing acoustic devices.

\begin{figure*}
    \centering
    \subfloat[]{\includegraphics[width=0.4\textwidth]{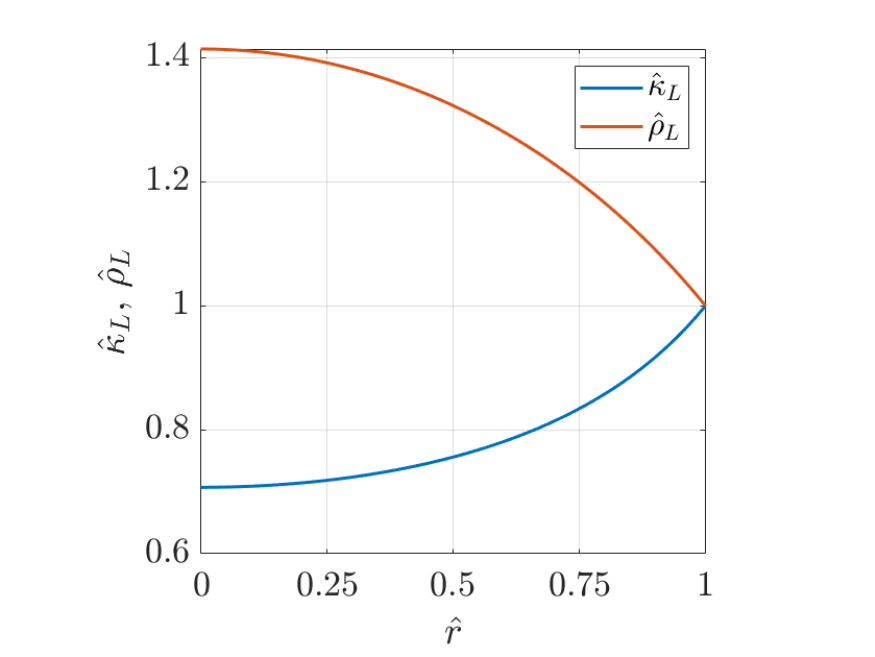}\label{fig:lens_settings-a}}
    \quad
    \subfloat[]{\includegraphics[width=0.27\textwidth,trim=0 -150 0 0]{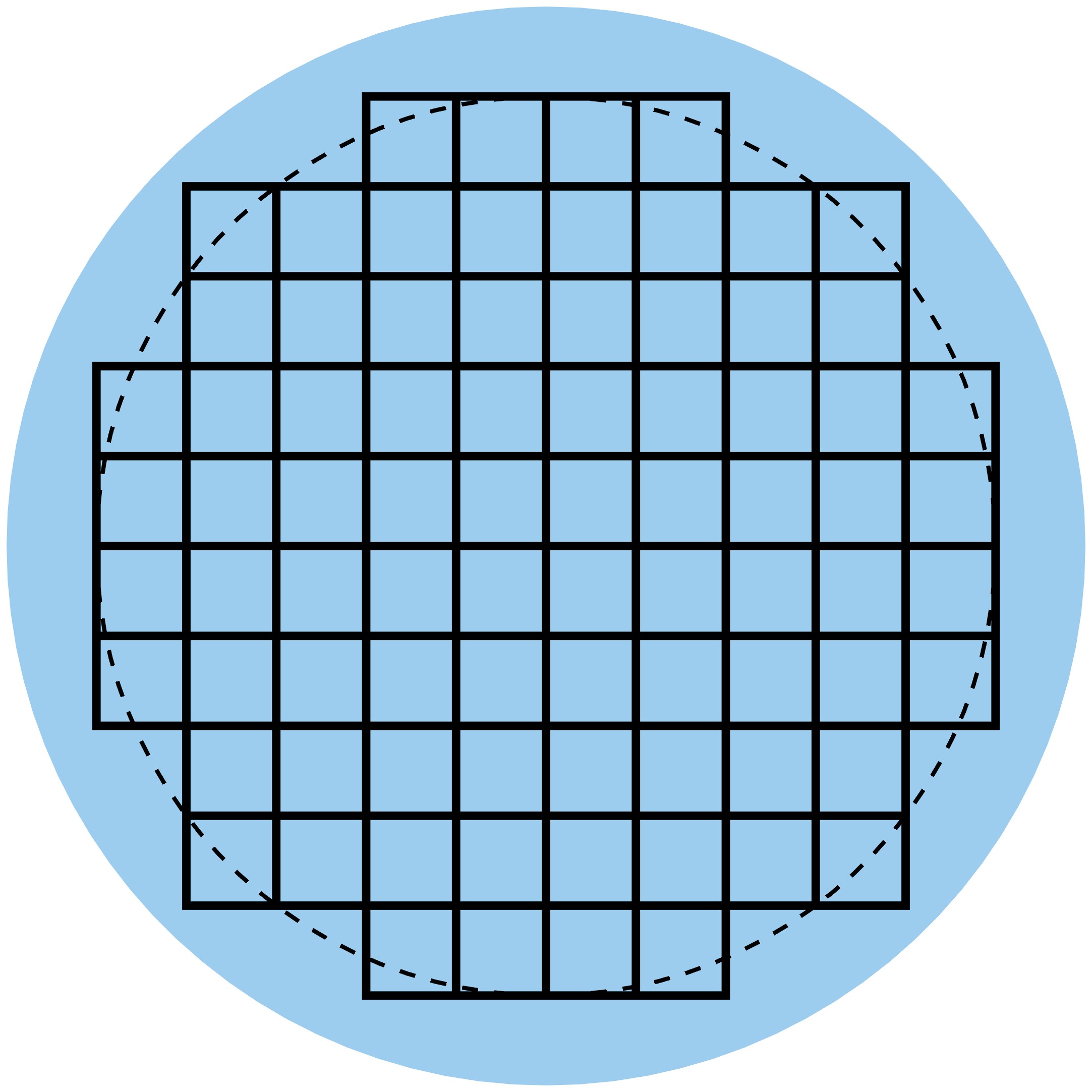}\label{fig:lens_settings-b} }
    \caption{(a) target properties of the lens normalized with respect to water, (b) ideal (dashed line) and discretized (solid line) volume of the lens.}
    \label{fig:lens_settings}
\end{figure*}

Developed within the geometric (ray) approximation of wave propagation, the L\"uneburg lens is defined by the following refractive index profile:
\begin{equation}
    \eta (\hat r) = \sqrt{2-\hat r^2},
\end{equation}
where $\hat r\coloneqq r/r_0$ is the non-dimensional radial coordinate and $r_0$ is the lens radius.
This specification determines the focusing behavior of the lens while leaving some freedom in the choice of the underlying acoustic properties \cite{cominelli2025non}.
We select them so that the impedance remains constant throughout the lens, considering it as a good compromise between transmission and realization. Thus:
\begin{align}\label{eq:acoustic_prop}
    \kappa_\mathrm{L}(\hat{r}) &= \frac{\kappa_0}{\eta(\hat r)},
    &
    \rho_\mathrm{L}(\hat{r}) &= \rho_0 \,\eta(\hat r),
\end{align}
where $\kappa_0 = \SI{2.2}{\giga\pascal}$ and $\rho_0 = \SI{1000}{\kilo\gram\per\cubic\meter}$ denote the bulk modulus and the density of water, respectively.
Fig.~\ref{fig:lens_settings-a} shows these property profiles in terms of non-dimensional properties $\hat\kappa_\mathrm{L}\coloneqq\kappa_\mathrm{L}/\kappa_0$ and $\hat\rho_\mathrm{L}\coloneqq\rho_\mathrm{L}/\rho_0$.

\subsection{Pentamode materials as artificial liquids}
In nature, there is no readily available fluid that supports acoustic waves while simultaneously exhibiting the properties required. Moreover, even if such fluids existed, maintaining them in the precise spatial configuration necessary for lens operation would be highly impractical. Pentamode Materials (PMs) \cite{milton1995which, norris2008acoustic} offer a promising alternative, as they can mimic the behavior of tailored fluids while providing the structural stability of a solid.

A PM is an elastic material that withstands Hook's law for solids, and the stress $\bs\sigma$ is related to the strain $\bs\varepsilon\coloneqq(\nabla\bb u + \nabla\bb u^\top)/2$ through the fourth-order elasticity tensor $\C$ as
\begin{equation}
    \bs\sigma = \C : \bs\varepsilon,
\end{equation}
where $:$ is the double contraction operator. Notably, a PM is such when $\C$ is (in good approximation) five times singular and can be written as
\begin{equation}\label{eq:C for PMs}
    \C = \kappa\,\bb S\otimes\bb S,
\end{equation}
where $\bb S$ is a second-order tensor and describes the only state of stress that the material supports; $\kappa$ is the bulk modulus, provided that $\bb S$ is normalized such that $\bb S:\bb S=1/3$; and $\otimes$ is the open tensor product. For a given strain $\bs\varepsilon$, the stress is proportional to $\bb S$
\begin{equation}
    \bs\sigma = (\kappa\bb S:\bs\varepsilon)\;\bb S
\end{equation}
and the state of stress can be described through the scalar value $p \coloneqq -\kappa\,\bb S:\bs\varepsilon$, also called pseudo pressure. Thus, the motion equation for the elastic material
\begin{equation}
    \nabla\cdot(\C:\nabla\bb u) = \rho\,\ddot{ \bb u}
\end{equation}
can be rearranged as a scalar equation for the pseudo-pressure:
\begin{equation}\label{eq:wave eq pseudo p}
    \kappa\,\bb S:\nabla(\rho^{-1}\bb S \nabla p) = \partial_{tt} p,
\end{equation}
provided that $\nabla\cdot\bb S=\bs0$ \cite{norris2008acoustic}. We indicate with $\partial_{tt}$ the second partial derivative with respect to time.
For the L\"uneburg lens, it is enough to consider an isotropic PM, having $\bb S$ equal to the identity tensor $\bb I$ so that Eq.~\eqref{eq:wave eq pseudo p} resembles the scalar wave equation for a heterogeneous isotropic still fluid. However, the anisotropy of PMs can be leveraged to obtain anisotropic fluids, as we do in Section~\ref{sec:cloak} to design an invisibility cloak.
In this work, we focus on the two-dimensional version of PMs, which are more properly called bimode materials, having only two easy modes of deformation in the plane we consider.

\subsection{Discretization and homogenization} \label{sec:homogenization}

Apart from liquids, the only PMs available in nature are gels, which pose a significant challenge for the construction of functional devices. However, recent advances in manufacturing technologies now enable the fabrication of engineered microstructures that closely approximate pentamode behavior \cite{Kadic2012,chen2017broadband,Quadrelli2021}. Importantly, for acoustic applications, it is not necessary to realize such materials at the atomic scale, but it suffices that the characteristic size of the microstructure is much smaller than the acoustic wavelength. This condition, known as the long-wavelength approximation, allows for a separation of scales. Under this regime, each unit cell can be treated as an effectively homogeneous medium, with its macroscopic elastic properties determined by solving the so-called \textit{cell problem}. For a detailed treatment of the homogenization framework, we refer the reader to \cite{Guedes1990, Laude2020}; here, we provide a brief overview in the context of the toy problem under consideration.

The circular domain of the lens is discretized into a rectangular grid consisting of 80 square cells, each measuring $\SI{20}{\milli\meter} \times \SI{20}{\milli\meter}$, as illustrated in Fig.~\ref{fig:lens_settings-b}. The grid size must be chosen based on the frequency range of interest: for homogenization to be valid, each cell must be much smaller than the acoustic wavelength.
As a proof of concept, we limit our analysis to 80 cells, and thus a discretization valid for a relatively low-frequency range will be obtained; however, all subsequent steps can be applied to finer discretizations as well.
More discussion about this is provided when a particular microstructure is analyzed in Section~\ref{sec:topopt}. We also note that a finer grid would provide a better approximation of the lens’s circular boundary.

\begin{figure*}
    \centering
    \subfloat[]{\includegraphics[width=0.3\textwidth]{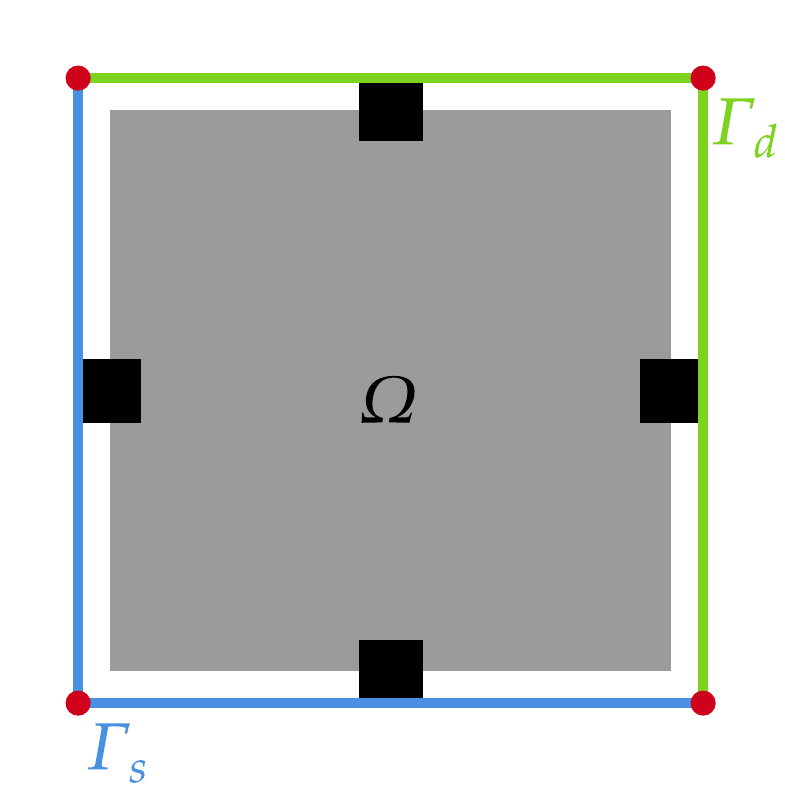}\label{fig:cell_settings-a}}
    \quad
    \subfloat[]{\includegraphics[width=0.3\textwidth]{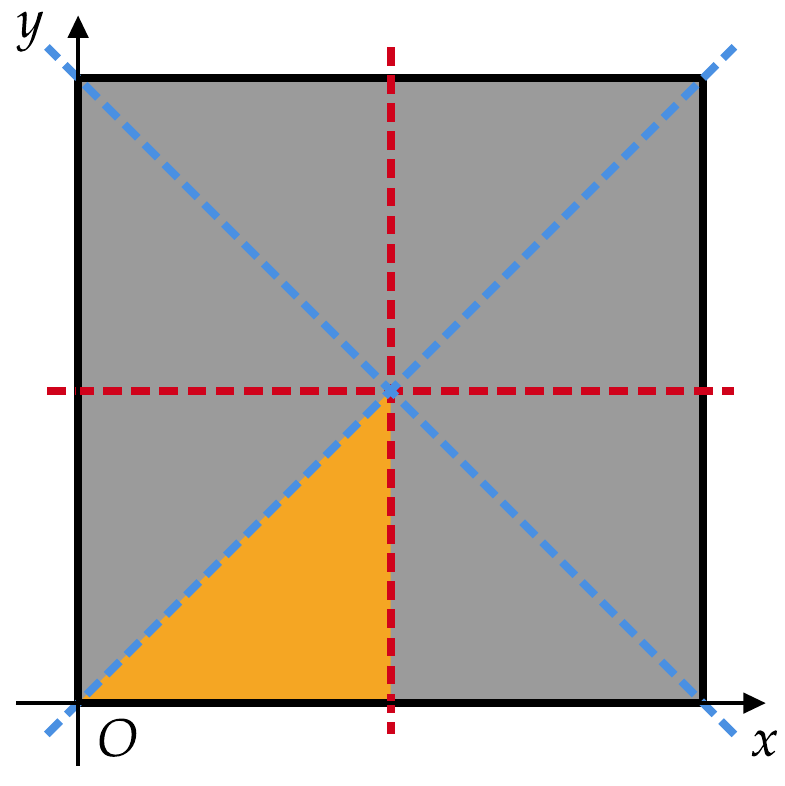}\label{fig:cell_settings-b}}
    \quad
    \subfloat[]{\includegraphics[width=0.3\textwidth]{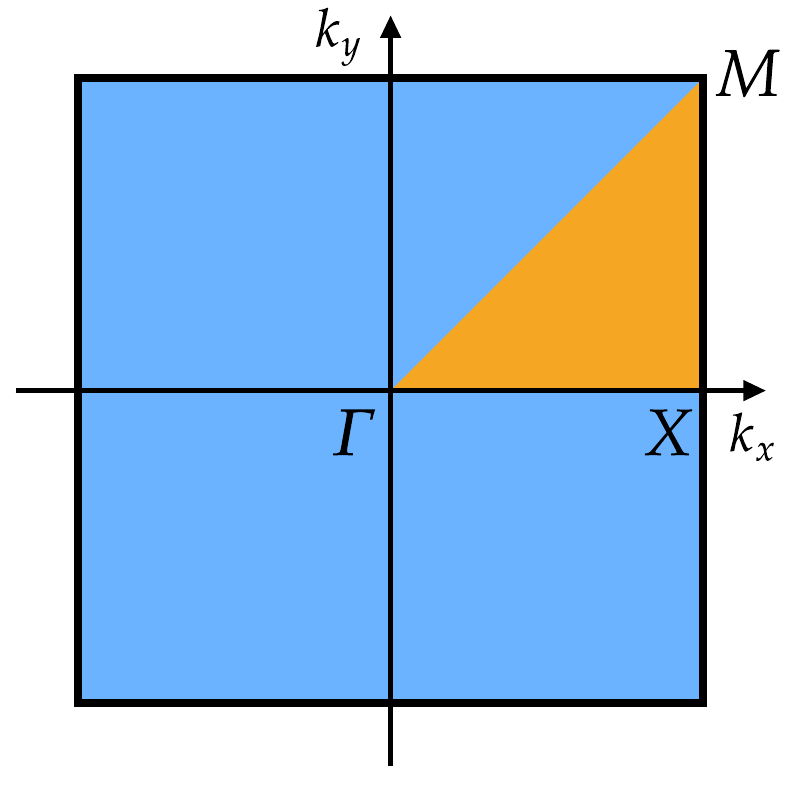}\label{fig:cell_settings-c}}    
    \caption{(a) The unit cell used as a first guess. The black and white regions stand for the connecting supports and the free frame, respectively. These regions remain fixed during the optimization routine to ensure connectivity between neighboring cells. The gray area is the design domain. The blue and green boundaries are the source and destination of the periodic conditions, respectively; the red points represent the node fixed to the ground to prevent rigid motion. (b) Symmetries imposed on the unit cell and the primitive design domain are highlighted in orange. (c) Unit cell of the reciprocal lattice with the irreducible Brillouin zone in orange.}
    \label{fig:cell_settings}
\end{figure*}

Each cell is considered independently using periodic boundary conditions, which treat the cell as if it were part of an infinite lattice. Although this approach ignores the lens global heterogeneity and finiteness, it is widely accepted in the literature as a valid approximation \cite{Laude2020}. This assumption becomes more accurate with finer discretization, as the structure becomes increasingly locally periodic.
\\
We consider the unit cell illustrated in Fig.~\ref{fig:cell_settings}, which contains a microstructure to be determined. An infinite periodic lattice is obtained by imposing that any field repeats periodically with respect to the cell domain $\Omega$, i.e.\ when passing from the blue edges $\Gamma_\mathrm{s}$ to the green ones $\Gamma_\mathrm{d}$.
Under the assumptions of low-frequency homogenization, the effective properties of the cell are determined using spatial averaging operators. The effective density is computed as the volume average of the local density, whereas the effective elastic properties require evaluating the cell response to distinct strain conditions.
In two dimensions, the averaged homogeneous stress $\bs\sigma^{\mathrm h}$ and strain $\bs\varepsilon^{\mathrm h}$ tensors have three independent components. Consequently, the linear mapping $\C^\mathrm{h} \colon \bs\varepsilon^{\mathrm h} \mapsto \bs\sigma^{\mathrm h}$ given by the homogenized tensor $\C^\mathrm{h}$ is characterized by applying three linearly independent strains $\bs\varepsilon_i^{\mathrm h}$, $i=1,2,3$. In 2D, these prescribed strains may be chosen as two uniaxial deformations and one shear deformation.
The effect is equivalent to applying three distinct anisotropic thermal expansions at every point of the cell.
The resulting strain tensors $\bs\varepsilon_i$, $\forall i = 1, 2, 3$, are computed by solving
\begin{subequations} \label{eq:static_equation}
\begin{align}
    &\nabla\cdot\bs\sigma_i = \nabla\cdot\big(\C:\bs\varepsilon^{\mathrm h}_i\big)
    \\
    &\bs\sigma_i=\C:\bs\varepsilon_i
    \\
    &\bs\varepsilon_i = \frac{\nabla \bb u_i + \nabla\bb u_i^\top}{2}
\end{align}
\end{subequations}
where the displacement $\bb u_i \in \mathcal{U} \subseteq [H^1(\Omega)]^2$ and the stress $\bs\sigma_i$ fields are $\Omega$-periodic and continuous in the cell domain and across its periodic boundaries $\Gamma_{\mathrm s}$ and $\Gamma_{\mathrm d}$. For the problem to be well-posed, we also ask that $\bs u(\bs0)=\bs 0$.
Using the FEniCSx library \cite{fenicsx}, the problem is efficiently implemented and solved in its weak form:
\begin{equation}
    \int_{\Omega}{ \left( \bs\varepsilon^{\mathrm h}_i - \bs\varepsilon_i \right) : \C : \nabla \bb v \,\mathrm dx} = 0, \quad \forall \bb v \in \mathcal{U}, \quad \forall i,j = 1, 2, 3,
\end{equation}
where $\bb v$ is the test function.
The homogenized elasticity tensor $\C^\mathrm{h}$ is computed component-wise as:
\begin{align} \label{eq:elasticity_tensor}
    & \C^\mathrm{h}_{ij} = \frac{1}{V_0} \int_{\Omega}{\left( \bs\varepsilon^{\mathrm h}_i - \bs\varepsilon_i \right) : \C : \left( \bs \varepsilon^{\mathrm h}_j - \bs\varepsilon_j \right) \,\mathrm dx},
    \quad \forall i,j = 1, 2, 3.
\end{align}

With the framework for analyzing individual cells and assembling the full lens now in place, the missing piece is the design of the microstructure itself. This is addressed by formulating and solving an inverse problem via topological optimization, as outlined in the next section.

\section{Inverse design through topology optimization}
\label{sec:topopt}

We now introduce the topology optimization framework used for the inverse design of the microstructure. We adopt the density-based formulation \cite{Bendsoe2004}, where the density field $\mu\in \mathcal M \subseteq L^2(\Omega)$ rescales the elastic properties of the cell problem \eqref{eq:static_equation} point-wise, such that $\C$ is replaced by $\mu\C$. The values of $\mu$ range from 0 (void) to 1 (solid), and the structure emerges as the subset of points where $\mu$ approaches~1. The function $\mu$ is found as the solution of the optimization problem defined in the following, but a set of filters is needed to regularize and stabilize the numerical solution.
\\
First, to avoid mesh-dependent solutions and the checkerboard problem \cite{Wang2011}, the filtered density field ${\tilde\mu\in \tilde{\mathcal{M}}\subseteq H^1(\Omega)}$ is computed as the solution of the Helmholtz equation \cite{lazarov2011filter}:
\begin{subequations} \label{eq:helmoltz_filter}
\begin{align}
    -R_f^2 \,\Delta \tilde\mu + \tilde\mu - \mu &= 0 \quad \text{in } \Omega,
    \\
    \nabla \tilde\mu \cdot \bs n &= 0 \quad \text{on } \Gamma,
\end{align}
\end{subequations}
where $R_f$ is the filter radius, selected as
\begin{equation}
    R_f = \frac{1}{2\sqrt{3}}\,\frac{l}{40}
\end{equation}
to apply a moving filter with a radius of approximately $1/40$ of the edge length $l$.
\\
Using FEniCSx \cite{fenicsx}, the filter is implemented in its weak form:
\begin{equation} \label{eq:filter_scheme}
    h(\mu, \tilde\mu, \tilde\psi) = \int_{\Omega} R_f^2 \, \nabla \tilde\mu \cdot \nabla \tilde\psi \, \mathrm{d}x +\int_{\Omega} (\tilde\mu - \mu) \, \tilde\psi \, \mathrm{d}x = 0, \quad \forall \tilde\psi \in \tilde{\mathcal{M}},
\end{equation}
where $\tilde\psi$ is the test function.
\\
The use of a regularization filter produces gray transition zones between solid and void regions, which can lead to inaccuracies in the modeling of physical behavior. Additionally, it also fails to prevent the appearance of one-node hinges in structures characterized by low stiffness values, e.g., compliant mechanisms \cite{Wang2011}. To address these limitations, the projection scheme proposed by \cite{Wang2011} is used:
\begin{equation} \label{eq:projection_scheme}
    \bar\mu = \frac{\tanh{( \beta \eta )} + \tanh{(\beta (\tilde\mu - \eta))}}{\tanh{(\beta \eta)} + \tanh{(\beta (1 - \eta))}},
\end{equation}
where $\bar\mu \in \tilde{\mathcal{M}}$ is the projected density, and $\beta = 10$ and $\eta = 0.5$ are projection parameters. The projected density field is used to compute the effective volume $V$ of the cell:
\begin{equation} \label{eq:volume}
    V = \int_{\Omega} \bar\mu\,\mathrm{d}x.
\end{equation}
Finally, the SIMP scheme \cite{Bendsoe1999} is used to obtain the physical densities $\hat{\mu}$ from the projected ones $\bar\mu$:
\begin{equation} \label{eq:simp_scheme}
    \hat{\mu} = \hat{\mu}_0 + (1 - \hat\mu_0)\, \bar\mu^p,
\end{equation}
where $p = 3$ is the penalization power used to further penalize the gray transition regions, and $\hat\mu_0 = 10^{-6}$ is an arbitrarily small density of the void element used to prevent singularities in the numerical method. The physical density $\hat\mu$ is the one used to interpolate the material properties and, consequently, the material elasticity tensor $\C$.

The PM used to build the lens must be isotropic and with zero shear resistance. Therefore, its elasticity tensor, written in Voigt notation, shall be
\begin{equation}
    \C_\mathrm{L} = \kappa_\mathrm{L}\,
    \begin{bmatrix}
        1 & 1 & 0 \\1 & 1 & 0 \\ 0 & 0 & 0
    \end{bmatrix},
\end{equation}
where $\kappa_\mathrm{L}$ is the target bulk modulus of the cell.
\\
To meet this requirement, we enforce the symmetries shown in Fig.~\ref{fig:cell_settings-b} in the cell. The red symmetry axes impose the lattice to be orthotropic and $\C_{13}^\mathrm{h} = \C_{23}^\mathrm{h} = \C_{31}^\mathrm{h} = \C_{32}^\mathrm{h} = 0$. The blue axes enforce symmetry with respect to $x$ and $y$, thus $\C_{11}^\mathrm{h} = \C_{22}^\mathrm{h}$.
We then formulate an optimization problem that minimizes the shear stiffness component $\C_{33}$ as follows:
\begin{equation} \label{opt:optimization_lens}
    \begin{aligned}
        \min_{\mu} \quad & \C^\mathrm{h}_{33} & \\
        \mathrm{s.t.} \quad & \C^\mathrm{h}_{11} \leq \kappa_\mathrm{L}(\hat r) & \\
        & \C^\mathrm{h}_{12} \geq 0.99\,\kappa_\mathrm{L}(\hat r) & \\
        & 0.99\,V_\mathrm{L}(\hat r) \leq V \leq V_\mathrm{L}(\hat r) & \\
        & 0 \leq \mu \leq 1
    \end{aligned}
\end{equation}
where $\kappa_\mathrm{L}$ and $V_\mathrm{L}$ are respectively the unit cell's target bulk modulus and effective volume that vary according to the radial position $\hat{r}$ measured from the center of the cell. Specifically, the value of $V_\mathrm{L}$ is computed from the density profile $\rho_\mathrm{L}$ considering the soft phase as void and the stiff phase composed of aluminum (Young's modulus $E_\mathrm{Al}=\SI{70}{\mega\pascal}$, Poisson's ratio $\nu_\mathrm{Al}=0.33$ and density $\rho_\mathrm{Al}=\SI{2700}{\kilo\gram\per\cubic\meter}$):
\begin{equation}
    V_\mathrm{L}(\hat r) = \frac{\rho_\mathrm{L}(\hat r)}{\rho_\mathrm{Al}}.
\end{equation}
Note that $\C_{11}^\mathrm{h}=\C_{22}^\mathrm{h}\ge\C_{12}^\mathrm{h}$ because $\C^\mathrm{h}$ is semi-positive definite. Therefore, by enforcing the first two constraints, the terms $\C_{11}^\mathrm{h}=\C_{22}^\mathrm{h}$ and $\C_{12}^\mathrm{h}$ are pushed to be close.
\\
Thanks to the lens symmetries, the full design space of $80$ cells is reduced to $11$ unique cells, each corresponding to a different radial position $\hat r$. Thus, 11 optimization problems~\eqref{opt:optimization_lens} are set independently and solved numerically: the cell domain is discretized into a structured grid of $200 \times 200$ square elements, and the design variable $\mu$ is approximated by a piecewise constant function over this grid. The optimization is then carried out using the Method of Moving Asymptotes (MMA) \cite{Svanberg1987}, a standard gradient-based optimization algorithm. This method requires the sensitivities of the objective and constraint functions, which are derived following the adjoint method (\ref{app:sensitivity}) and implemented using FEniCSx \cite{fenicsx}.

\begin{figure*}
    \centering
    \subfloat[]{\includegraphics[width=0.45\textwidth]{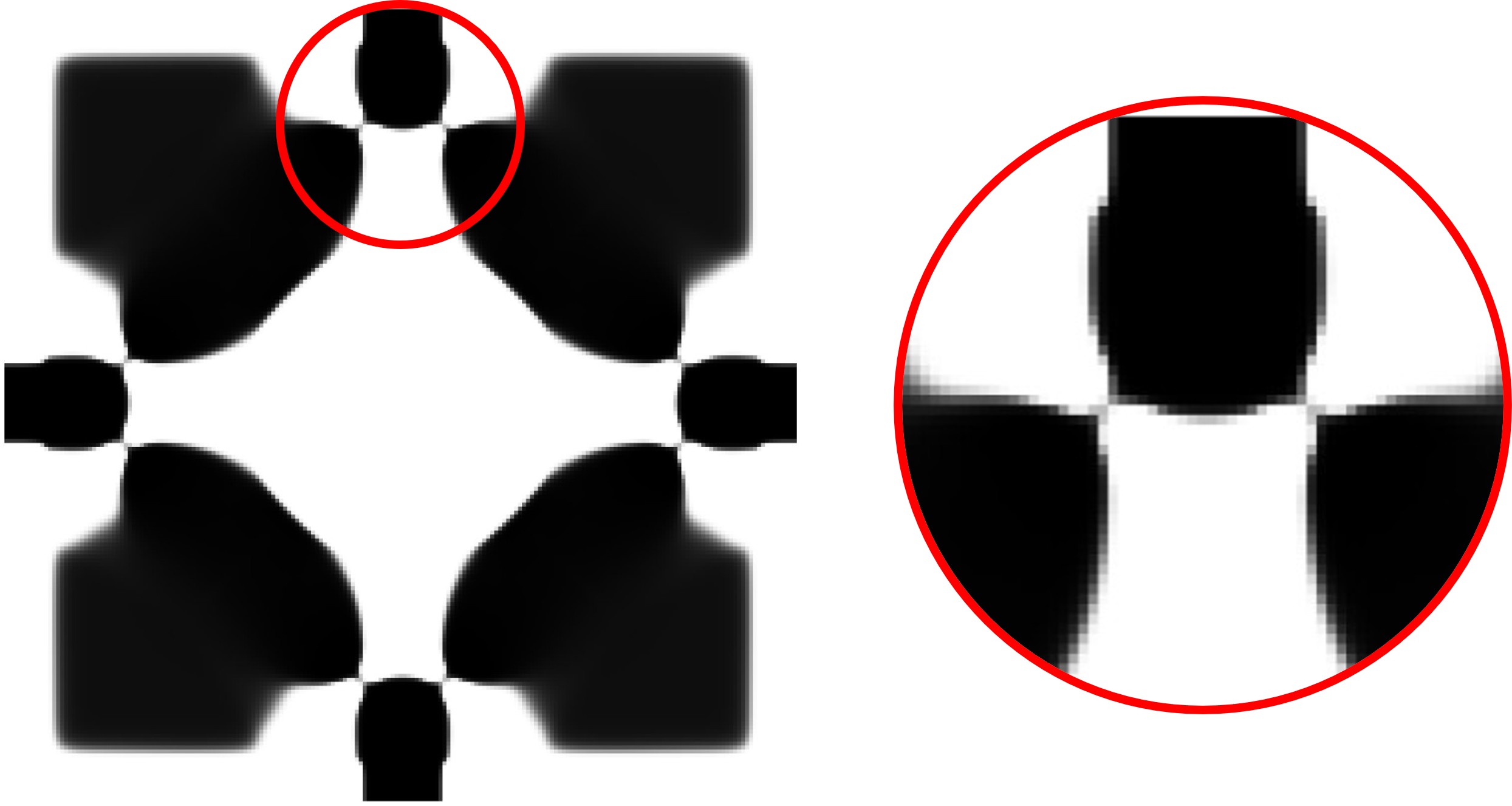}\label{fig:cell_issue}}
    \quad
    \subfloat[]{\includegraphics[width=0.45\textwidth]{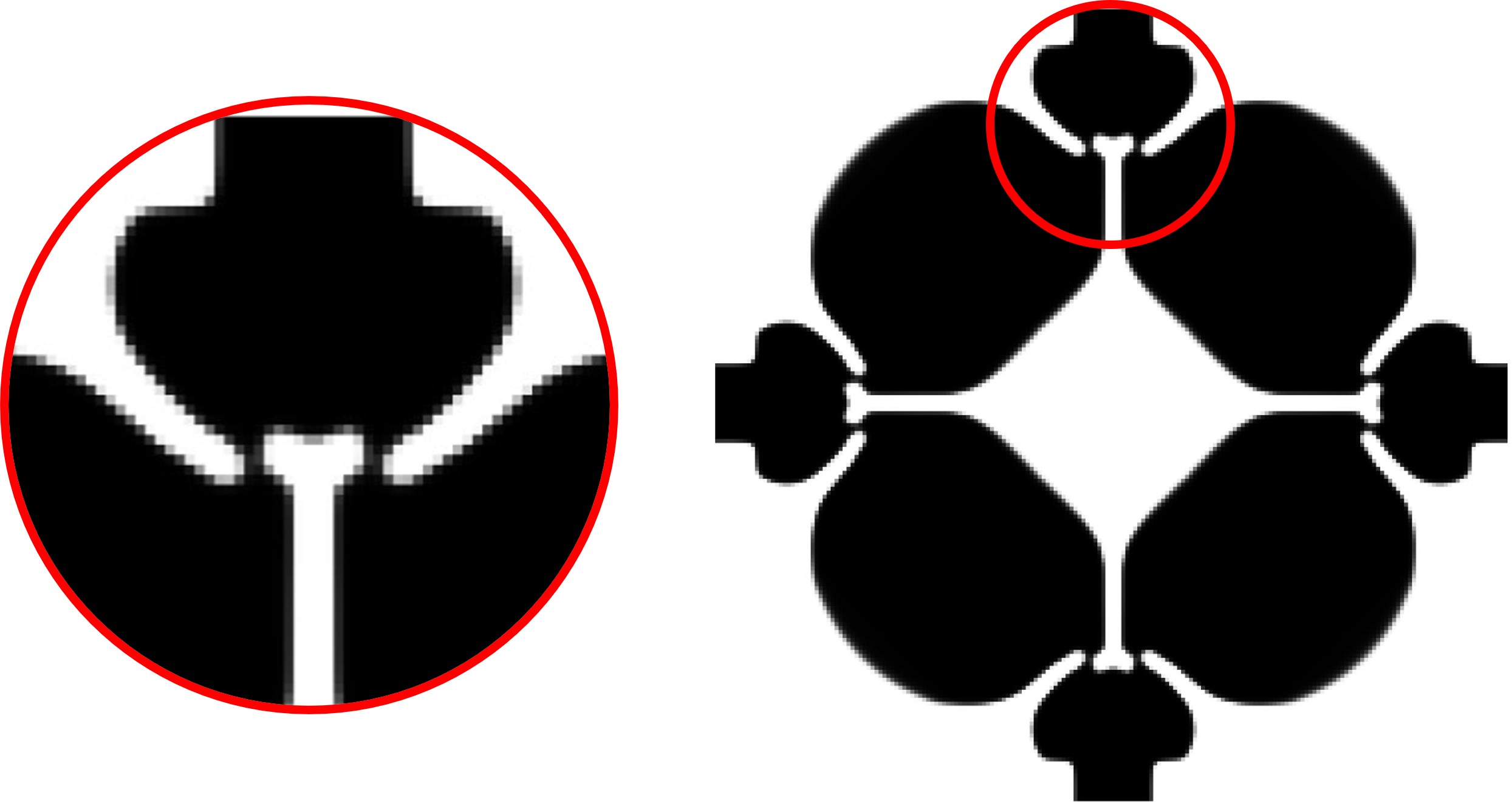}\label{fig:cell_vtm}}
    \caption{(a) Connectivity issue of the optimized cell obtained solving problem~\eqref{opt:optimization_lens} with $\hat{r} = 0.14$. The layout has non-physical connections that cannot be manufactured. (b) Cell obtained solving problem~\eqref{opt:optimization_lens_vtm} using the Virtual Temperature Method (VTM). The layout has no connectivity issues.}
    \label{fig:cell_issue_vtm}
\end{figure*}

A representative geometry obtained by solving problem~\eqref{opt:optimization_lens} is shown in Fig.~\ref{fig:cell_issue}.
The cell topology resembles the hexagonal lattice typically adopted for bimode materials \cite{chen2017broadband,Quadrelli2021}, where stiff massive links are connected by small elements that approximate hinges.
A nearly ideal pentamode having the desired mechanical properties is achieved,  with $\mathbb{C}_{33}^\mathrm{h} \approx 0$. This performance is likely facilitated by the smooth density transitions, especially around the functional hinges, which help minimize shear stiffness.
However, these same features pose significant challenges for fabrication as the density distribution lacks well-defined contours, making the realization impractical.

\subsection{Connectivity constraint} \label{sec:connectivity}

In the topology optimization field, connectivity constraints control the presence of enclosed holes or solid islands \cite{Cool2025connectivity, Bonaccorsi2025connectivity}. Following a similar approach as in \cite{Swartz2022periodic}, the Virtual Temperature Method (VTM) \cite{Liu2015, Li2016} is used to deal with the connectivity issue shown in Fig~\ref{fig:cell_issue}. The basic idea is to set up a thermal conduction problem described by:
\begin{subequations}\label{eq:thermal_conduction}
\begin{align} 
    \hat\mu \,\kappa_\mathrm{T} \Delta T &= 0,\phantom Q \quad \text{in } \Omega
    \\
    T &= 0,\phantom Q \quad \text{on } \Gamma_\mathrm{HS}
    \\
    \nabla T\cdot \bs n  &= Q,\phantom 0 \quad \text{on } \Gamma_Q
    \\
    \nabla T\cdot \bs n &= 0,\phantom Q \quad \text{on } \Gamma\setminus(\Gamma_Q\cup\Gamma_\mathrm{HS})
\end{align}
\end{subequations}
where $T$ is the temperature field, $\kappa_{\mathrm{T}}$ is the thermal conductivity, and $Q$ is a constant heat generation applied on $\Gamma_Q$. In this work, $\kappa_{\mathrm{T}}$ and $Q$ are both set to $1$. The homogeneous Dirichlet boundary condition is used to set a \textit{heat-sink} on $\Gamma_\mathrm{HS}$. The remaining portion of the boundary $\Gamma\setminus(\Gamma_Q\cup\Gamma_\mathrm{HS})$ is set adiabatic. As a result, by placing $\Gamma_\mathrm{HS}$ and $\Gamma_Q$ on two opposite sides (e.g., left and right), the heat generated on $\Gamma_Q$ flows through the structure towards $\Gamma_\mathrm{HS}$, thus inducing a heat flux that can be enforced to promote connectivity.
Using again FEniCSx \cite{fenicsx}, the thermal conduction problem is implemented and solved using its weak form:
\begin{equation}\label{eq:thermal_conduction_weak}
    \int_{\Omega} \hat\mu \,\kappa_{\mathrm{T}} \nabla T \cdot \nabla \tau \,\mathrm{d}x - \int_{\Gamma_Q} Q \tau \,\mathrm{d}s = 0, \quad \forall \tau \in \mathcal{T} \subseteq H^1(\Omega),
\end{equation}
where $\tau$ is the test function.
\\
After solving Eq.~\eqref{eq:thermal_conduction_weak}, the thermal compliance $C_{\mathrm{T}}$ is defined as
\begin{equation}
    C_{\mathrm{T}} = \int_{\Gamma_Q} Q\,T\,\mathrm dx.
\end{equation}
In line with the optimization algorithm, the sensitivity expressions of $C_{\mathrm{T}}$ are computed as shown in \ref{app:sensitivity}.

The cell connectivity is promoted by minimizing $C_{\mathrm{T}}$ \cite{Li2016}. Therefore, the optimization problem \eqref{opt:optimization_lens} is modified as follows:
\begin{equation} \label{opt:optimization_lens_vtm}
    \begin{aligned}
        \min_{\mu} \quad & C_{\mathrm{T}} & \\
        \mathrm{s.t.} \quad & \C^\mathrm{h}_{11} \leq \kappa_\mathrm{L}(\hat r) & \\
        & \C^\mathrm{h}_{12} \geq 0.99\,\kappa_\mathrm{L}(\hat r) & \\
        & \C^\mathrm{h}_{33} \leq 0.01\,\kappa_\mathrm{L}(\hat{r}) & \\
        & 0.99\,V_\mathrm{L}(\hat r) \leq V \leq V_\mathrm{L}(\hat r) & \\
        & 0 \leq \mu \leq 1,
    \end{aligned}
\end{equation}
where the shear modulus $\C_{33}^\mathrm{h}$ is now controlled with a constraint. The cell obtained with $\hat{r} = 0.14$ is shown in Fig.~\ref{fig:cell_vtm}. Although the geometry of this solution is similar to the previous one, shown in Fig.~\ref{fig:cell_issue}, the new cell presents well-defined hinge connections. Additionally, the new geometry appears to be better defined, with a clear boundary between the solid and void regions.

The constraint on the shear modulus $\C^\mathrm{h}_{33}$ is dictated by the target application. The authors have experienced that adjusting this constraint effectively regulates the thickness of the connecting features and can indirectly enforce a minimum length scale within the microstructure.

We now have all the tools to optimize the cells and build the lens microstructure.

\section{L\"uneburg lens} \label{sec:lens}

\begin{figure*}
    \centering
    \includegraphics[width=0.8\textwidth]{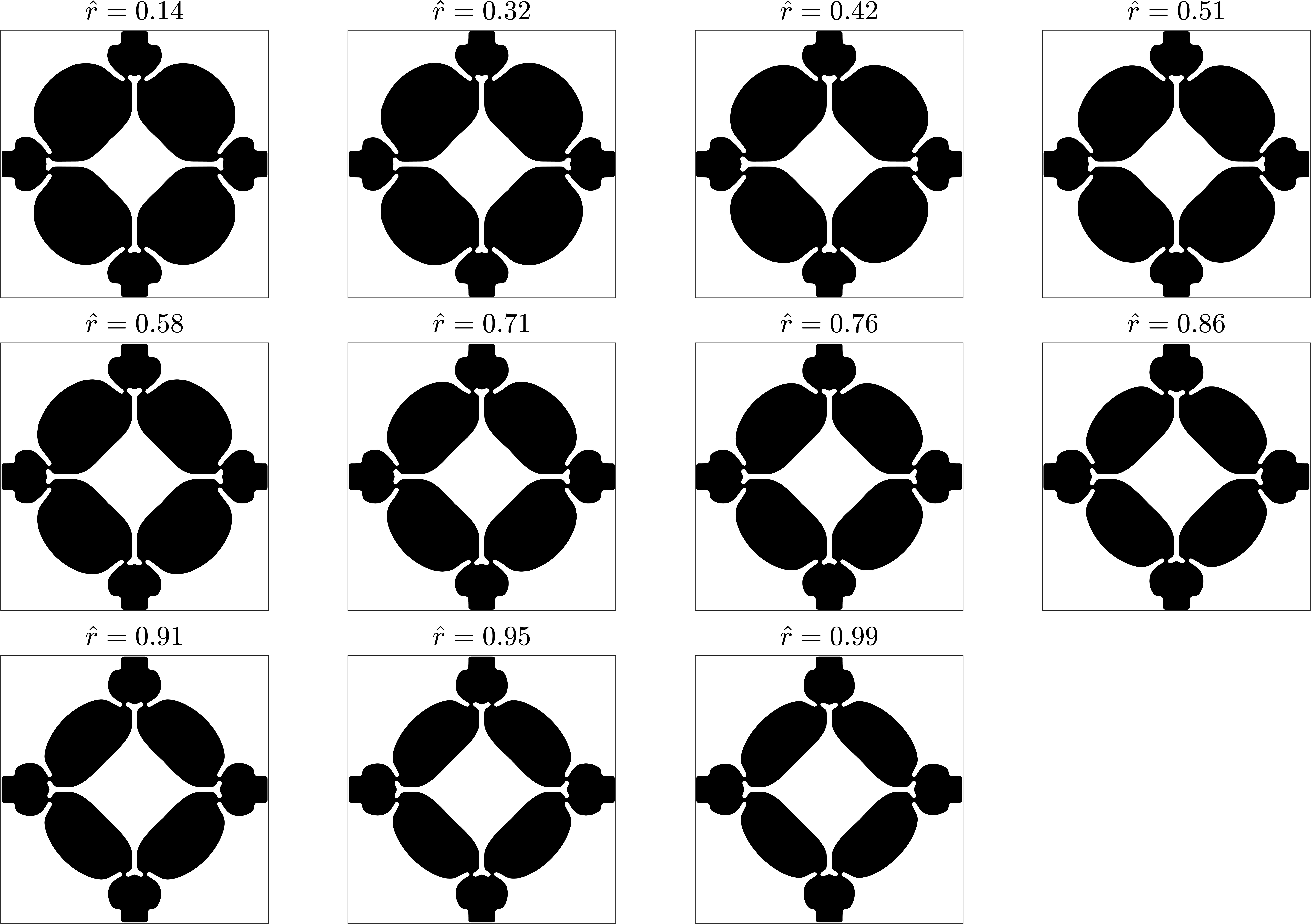}
    \caption{Optimized lens cells obtained solving problem \eqref{opt:optimization_lens_vtm}.}
    \label{fig:lens_cells}
\end{figure*}

\begin{figure*}
    \centering
    \subfloat[]{    \includegraphics[height=0.3\textwidth,trim=25 0 25 0,clip]{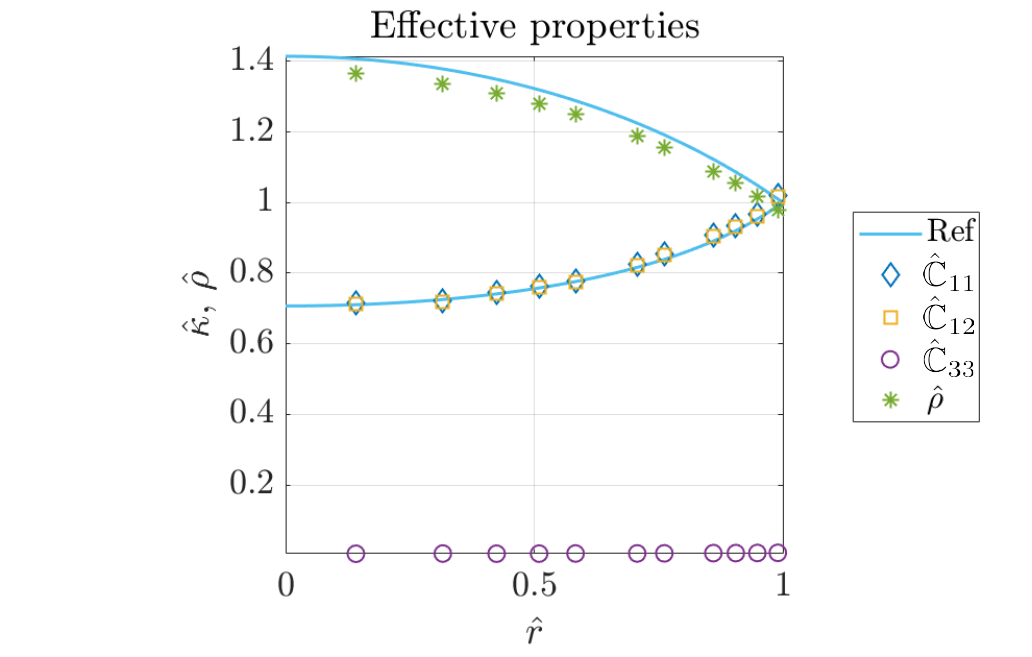}\label{fig:lens_properties-a}}
    \subfloat[]{    \includegraphics[height=0.3\textwidth,trim=25 0 25 0,clip]{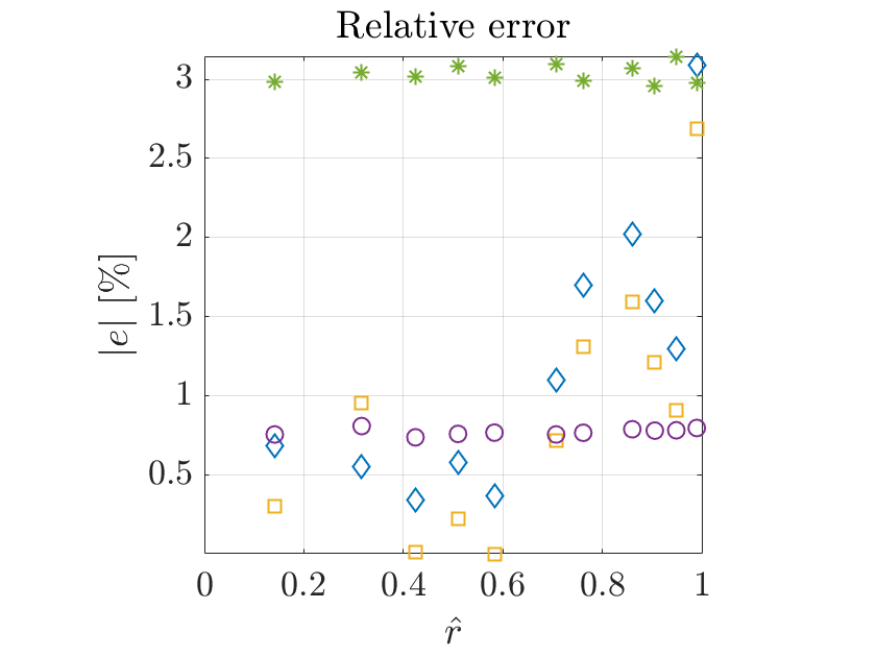}\label{fig:lens_properties-b}}
    \caption{Left, the effective properties (normalized with respect to water) of the optimized cells shown in Fig.~\ref{fig:lens_cells}, computed using \Comsol{}. Right, the magnitude of the relative error between effective and target properties.
    }
    \label{fig:lens_properties}
\end{figure*}

The optimization problem~\eqref{opt:optimization_lens_vtm} is solved using a workstation equipped with Intel Core i9-13900KS CPU at \SI{3.20}{\giga\hertz} and \SI{128.0}{\giga\byte} RAM. Using parallel computing with distributed memory via MPI \cite{walker1996mpi} over four processors, the optimization of each cell completed 100 iterations--enough to reach convergence--in approximately 1 minute (0.6 seconds per iteration), resulting in an overall computational time of about 11 minutes for the 11 cells.
\\
The density distributions obtained for the 11 optimized cells are shown in Fig.~\ref{fig:lens_cells}. Each design satisfies the target properties with a relative tolerance of \SI{1}{\percent}, as required by the optimization problem~\eqref{opt:optimization_lens_vtm}. The geometry contours are obtained from the optimal densities $\mu$ using the \texttt{contourc} \Matlab{} function with a threshold of $0.5$, and every cell is simulated using \Comsol{} to evaluate its effective properties with a body-fitted mesh. Due to the fixed-grid approximation inherent in the topology optimization framework, some discrepancies with the \Comsol{} simulations are expected. Despite this, the comparative plot shown in Fig.~\ref{fig:lens_properties} reveals that the maximum relative error between the simulated and target properties is approximately \SI{3}{\percent}, which is considered acceptable for the proposed application.

\subsection{Lens analysis} \label{sec:lens_analysis}

\begin{figure*}
    \centering
    \subfloat[]{\includegraphics[height=0.25\textwidth,trim=33 15 100 0,clip]{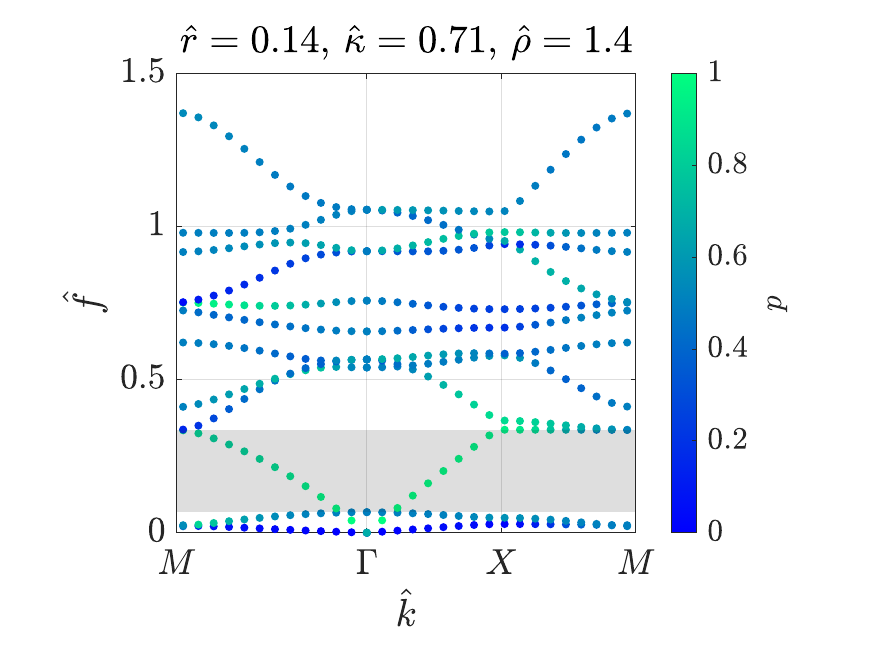}\label{fig:lens_dispersion-a}}
    \subfloat[]{\includegraphics[height=0.25\textwidth,trim=50 15 100 0,clip]{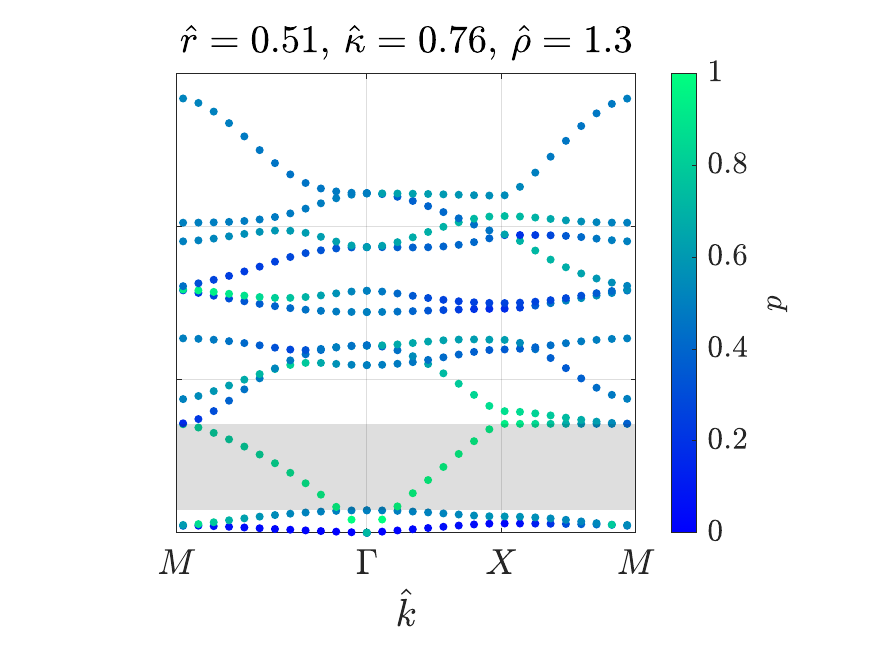}\label{fig:lens_dispersion-b}}
    \subfloat[]{\includegraphics[height=0.25\textwidth,trim=50 15 55 0,clip]{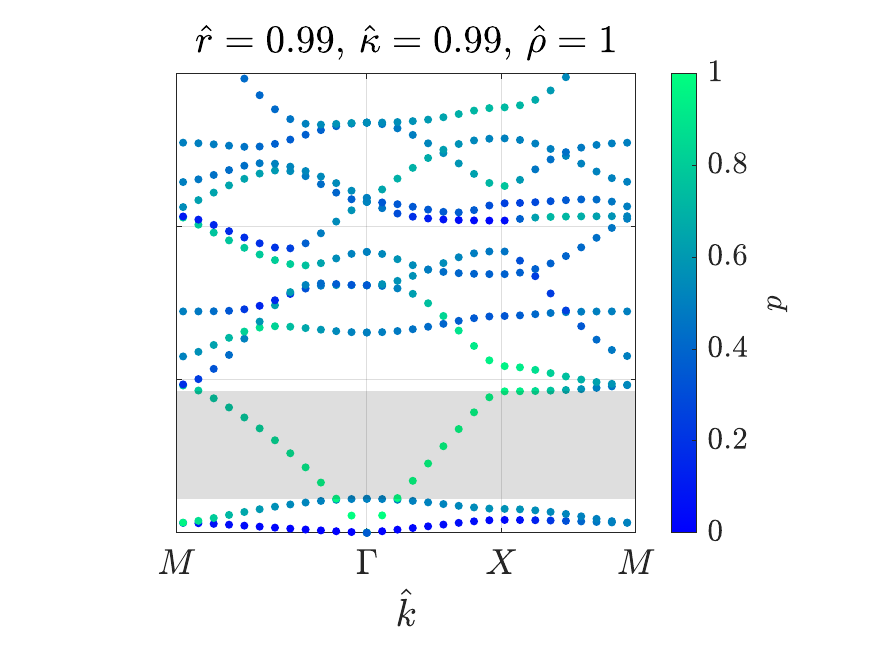}\label{fig:lens_dispersion-c}}
    \subfloat[]{\includegraphics[height=0.24\textwidth,trim=10 5 0 0,clip]{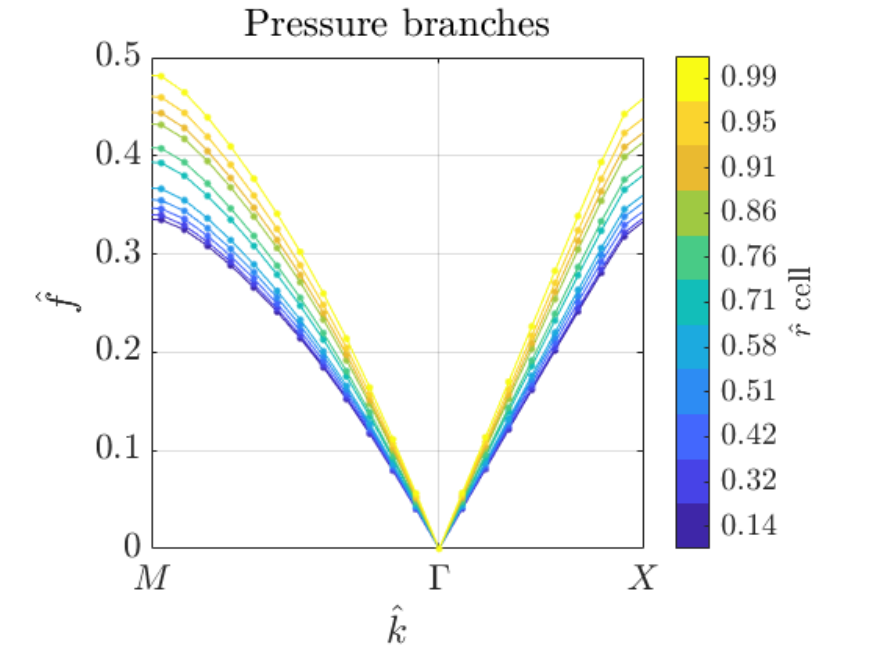}\label{fig:lens_dispersion-d}}
    \caption{(a)-(c) Dispersion diagrams of three representative optimized cells of Fig.~\ref{fig:lens_cells}, computed using \Comsol{}. The pentamode bandgaps are highlighted in gray.
    (d) Summary of the longitudinally-polarized acoustic branches.
    }
    \label{fig:lens_dispersion}
\end{figure*}

The effective properties in the long-wavelength limit are computed in the quasi-static regime, so a verification of the dynamic properties of the lattice is needed.
We compute the dispersion diagram of each cell through \Comsol{}. Fig.~\ref{fig:lens_dispersion} shows the dispersion branches of three representative cells, including the innermost and outermost cells. We limit ourselves to the path $\Gamma XM \Gamma$ that surrounds the Irreducible Brillouin Zone (IBZ) shown in Fig.~\ref{fig:cell_settings-c}, obtained using the symmetries of a square cell. The frequency is adimensionalized with respect to the cell edge $l$ and the water sound speed $c_0=\sqrt{\kappa_0/\rho_0}$ as $\hat f \coloneqq f\; l/c_0$.
The dispersion branches have a color scale ranging from 0 to 1 and indicates the polarization $\mathcal P$ of the modes defined as
\begin{equation}\label{eq:polariz}
    \mathcal P\coloneqq \frac{\int_\Omega (\bb k\cdot\bb u)\,\overline{(\bb k\cdot\bb u)} \,\mathrm d\Omega}{\int_\Omega \bb u\cdot \overline{\bb u}  \,\mathrm d\Omega},
\end{equation}
where $\bb u$ is the complex-valued displacement obtained by the eigenvalue problem use to compute the dispersion; a variable with a bar above it represents its complex conjugate.
A polarization close to 1 or 0 indicate a longitudinal or transversal wave, respectively.
In particular, we note that at $\hat f\in[0.11, 0.34]$ all the dispersion diagrams have the so-called shear bandgap typical of PMs, a frequency range where shear waves experience spatial attenuation while pressure waves are efficiently transmitted.
\\
Fig.~\ref{fig:lens_dispersion-d} collects the longitudinally-polarized acoustic branches of the 11 cells. In the low-frequency limit, these branches can be approximated as straight lines with different slopes due to the change in phase velocity. 
As the frequency increases, dispersive behavior arises.

\begin{figure*}
	\centering
    \begin{subfigure}{.32\textwidth}
        \centering
        \includegraphics[width=\textwidth,trim=55 10 45 20,clip]{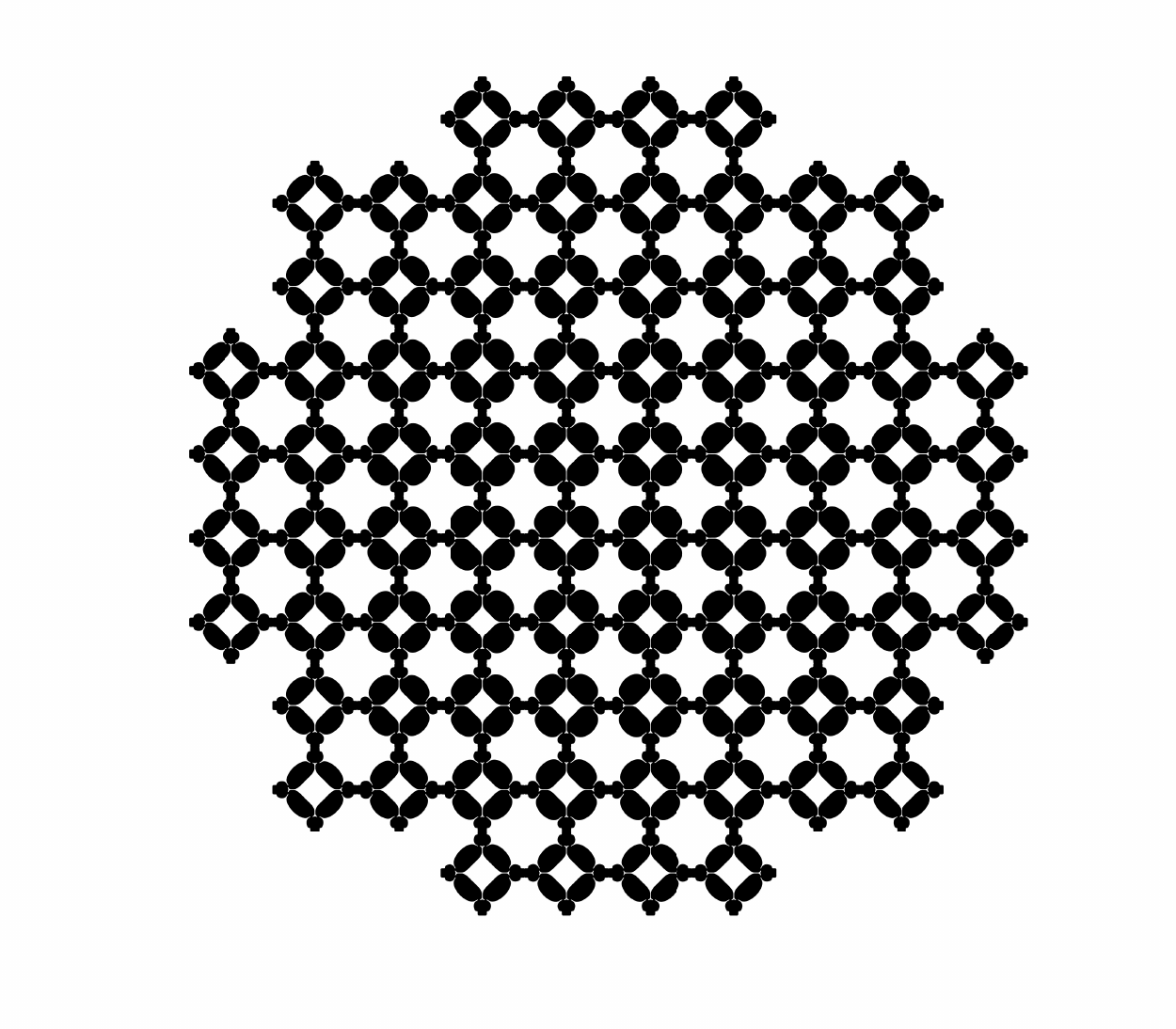}
        \caption{}\label{fig:lens_pressure-a}
    \end{subfigure}
    \begin{subfigure}{.33\textwidth}
    \centering
    \begin{tikzpicture}
        \matrix[matrix of nodes,
                row sep=-1pt, column sep=-1pt,
                nodes={inner sep=0, anchor=center}] (main) {
            \includegraphics[width=0.49\textwidth,trim=70 25 70 0,clip]{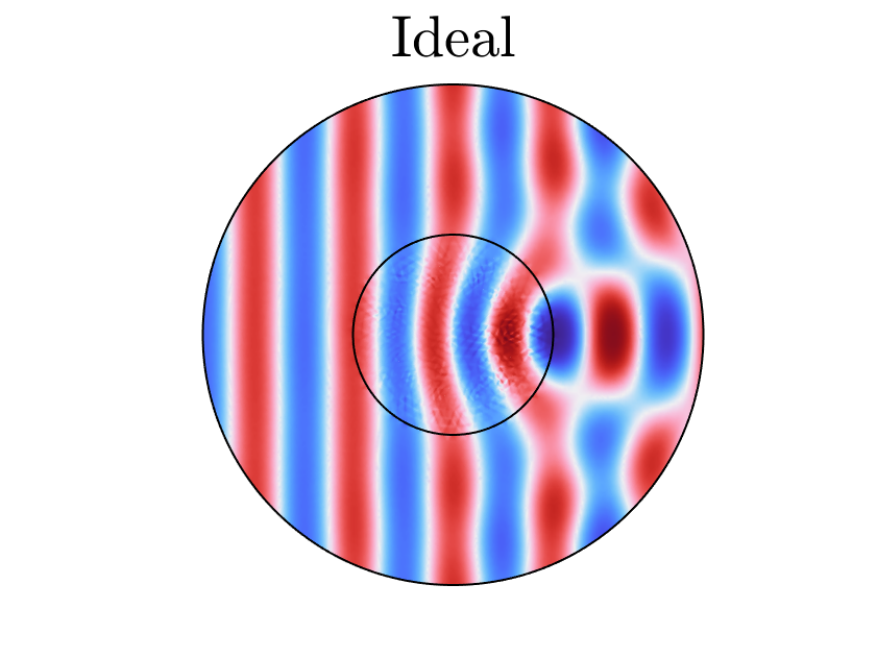} &
            \includegraphics[width=0.49\textwidth,trim=70 25 70 0,clip]{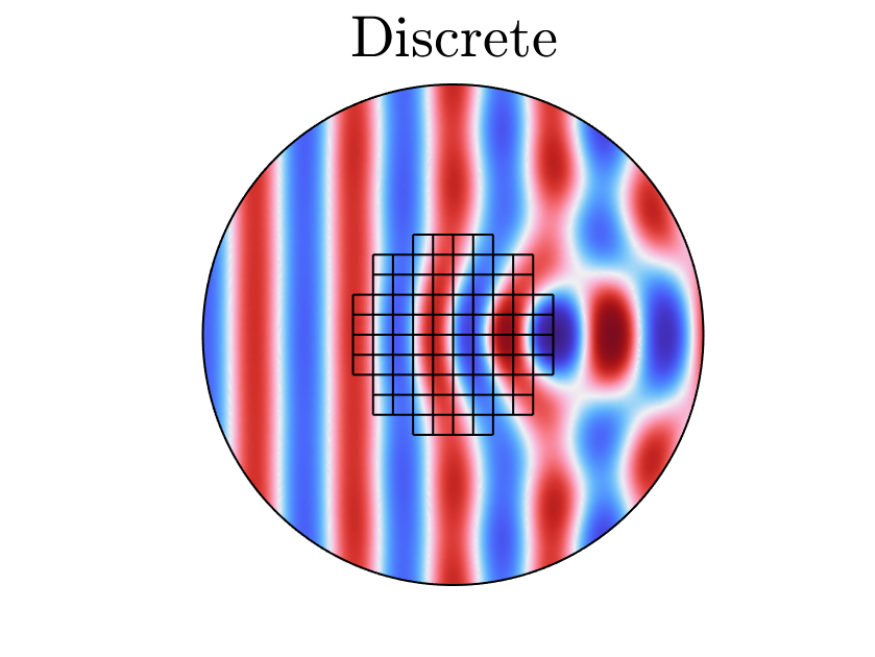} \\
            \includegraphics[width=0.49\textwidth,trim=70 25 70 0,clip]{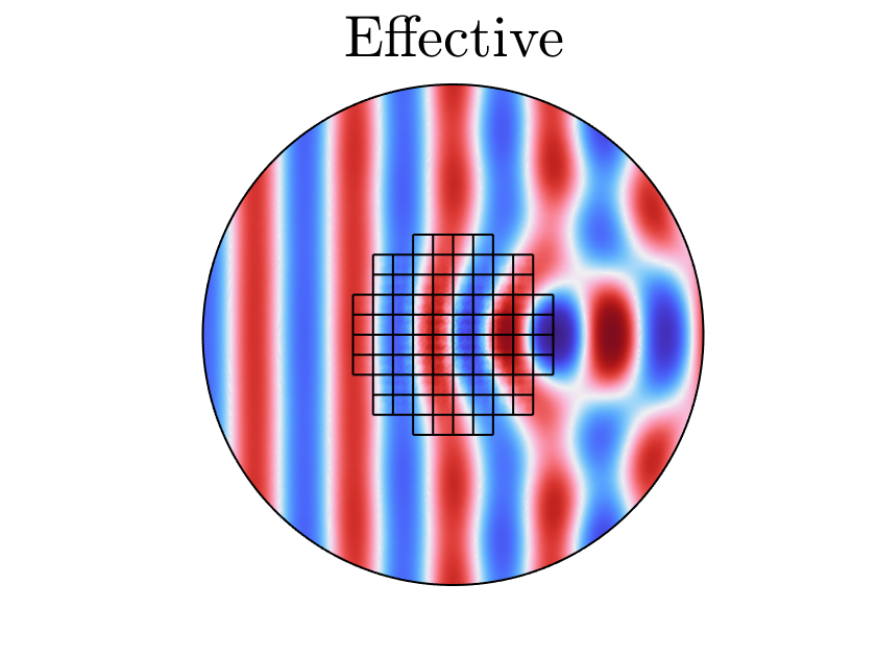} &
            \includegraphics[width=0.49\textwidth,trim=70 25 70 0,clip]{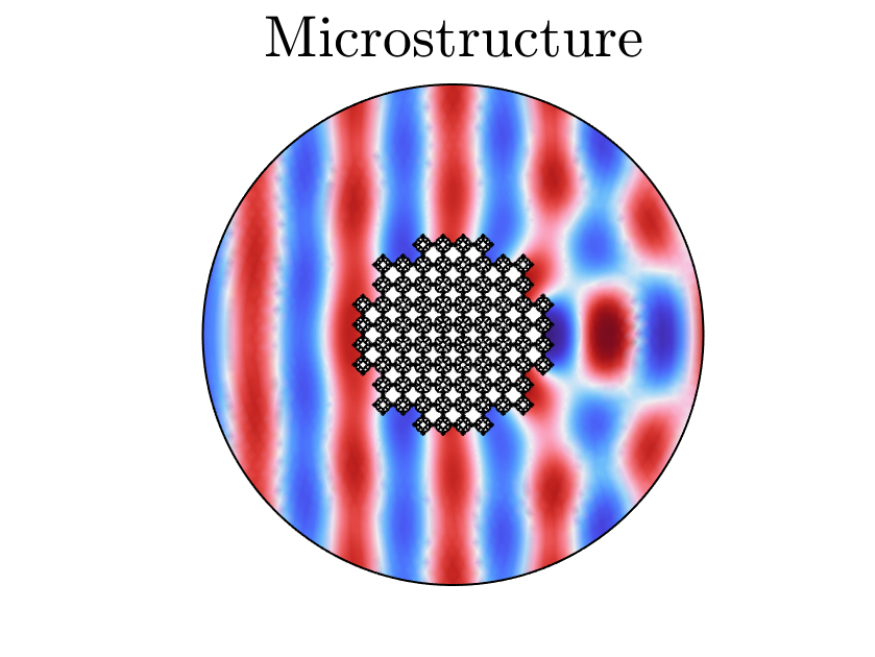} \\
        };
        \node[anchor=center] at (main.center)
            {\includegraphics[width=0.12\textwidth,trim=-20 0 0 -20]{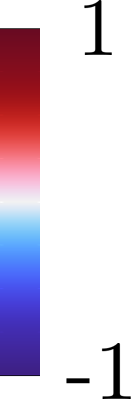}};
    \end{tikzpicture}
    \caption{}
    \label{fig:lens_pressure-b}
    \end{subfigure}
    \begin{subfigure}{.3\textwidth}
        \centering
        \includegraphics[width=\textwidth,trim=50 0 50 0,clip]{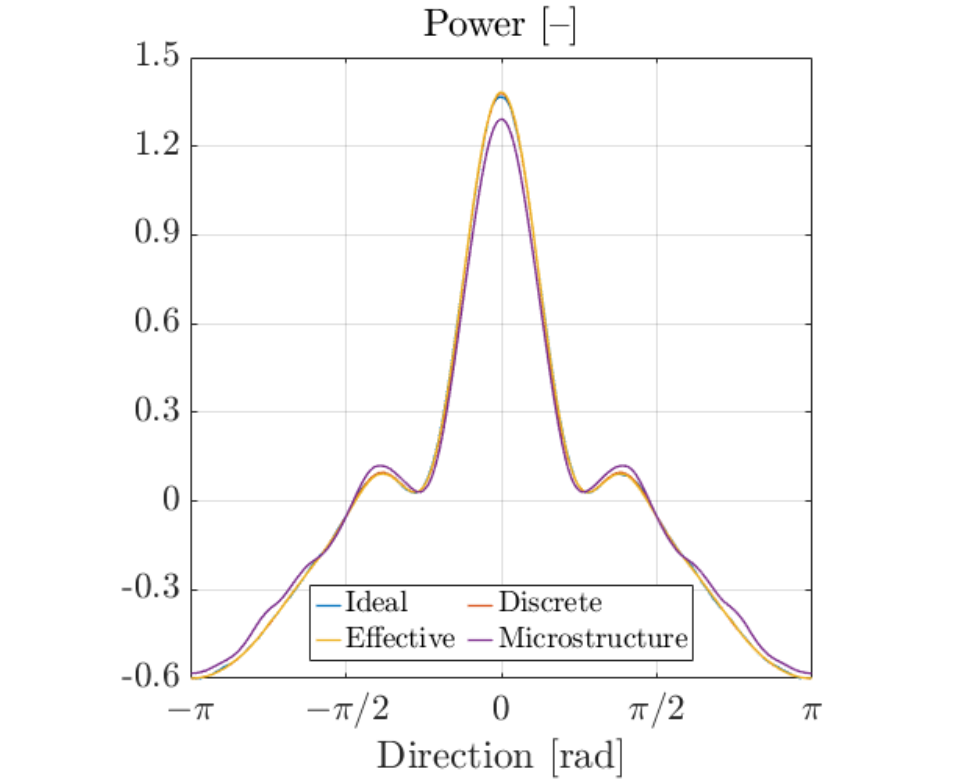}
        \caption{}\label{fig:lens_pressure-c}
    \end{subfigure}    
	\caption{(a) microstructure of the lens; (b) real part of the pressure field phasor, arbitrary units. In the elastic domains of the discrete and the effective lens, the pseudo pressure is displayed; (c) power directivity.
	}
	\label{fig:lens_pressure}
\end{figure*}

Finally, the lens is constructed by repeating the 11 primitive cells according to their symmetries.
The microstructure is simulated in \Comsol{} for $\hat f=0.2$, where a background plane wave propagating from the left is imposed in a circular domain truncated by Perfectly Matched Layers (PMLs). Fig.~\ref{fig:lens_pressure-b} shows the pressure field for four consecutive approximation steps: an ideal lens with continuous properties given by Eq.~\eqref{eq:acoustic_prop}, the discretized lens with ideal properties, the discretized lens with optimized effective properties, and the full elastic microstructure. In all scenarios, the plane wave is focused on the right-hand side. The directivity plot in Fig.~\ref{fig:lens_pressure-c} provides a quantitative comparison of the non-dimensional power, $\hat P\coloneqq P\;\rho_0 c_0/(p^2rh)$, crossing a cylindrical angular section of radius $r$ and height $h$, for the four lenses. This shows that there is good agreement between them.

\section{Acoustic cloak} \label{sec:cloak}

In this section, we revisit the framework previously applied to the L\"uneburg lens and extend it to the more challenging case of acoustic cloaking, thereby demonstrating the versatility of our approach. In this context, the required PM must exhibit not only heterogeneity but also anisotropy. We focus on the canonical benchmark of a circular target to be cloaked—an extensively studied configuration in the literature and one for which experimental validation was provided by Chen \textit{et al.}~\cite{chen2017broadband}. In their work, the unit cell geometry was defined by a set of parameters, requiring careful design and parametric optimization. Thanks to axial symmetry, the number of independent cells was limited, and many could be repeated around the circumference. In contrast, a large number of distinct cells is required for more intricate geometries, and our method would allow their design efficiently and without relying on predefined parameterizations.

We shortly summarize the results of transformation acoustics (TA) \cite{cummer2007one}, needed to define the properties of an invisibility cloak. For more details, the reader is referred to the comprehensive description by Norris \cite{norris2008acoustic} and similar implementations in the literature \cite{chen2017broadband,Quadrelli2021}. Let us consider two annular circular domains $\Omega$ and $\omega$ with internal radii $\delta$ and $a>\delta$, respectively, and the same external radius $b>a$.
According to TA, the deformation induced by a map $\bs\chi\colon \bb X\in\Omega \mapsto\bb x\in\omega$ can be reinterpreted as a modification of the acoustic properties. In particular, we consider a domain $\Omega$ filled by pure water (i.e.\ $\rho=\rho_0$ and $\kappa=\kappa_0$) and the linear transformation $\bs\chi$ defined in polar coordinates as
\begin{equation}
	\begin{dcases}
		r = \big(R-\delta\big)\frac{b-a}{b-\delta}+a \\
		\theta = \Theta
	\end{dcases},
\end{equation}
where capital and lowercase letters refer to the virtual $\Omega$ and physical $\omega$ domains, respectively.
The elastic properties for a pure pentamode cloak are \cite{norris2008acoustic}
\begin{align}\label{eq:wave eq spatial}
    \rho =\rho_0\, J^{-1},  \qquad \C = \kappa_0\,J^{-1}\,\bb V\otimes\bb V,
\end{align}
where $\bb F \coloneqq\frac{\partial\bb x}{\partial\bb X}$ is the deformation gradient of $\bs\chi$, $J\coloneqq\det\bb F$ its determinant, and $\bb V$ the left stretch tensor (i.e.\ $\bb V^2=\bb F \bb F^\top$).
\\
The resulting effect is that the pressure field around the small obstacle of radius $\delta$ is deformed to fit around a larger obstacle of radius $a$, while the field beyond radius $b$ remains unaffected. Consequently, the two configurations are indistinguishable to a distant observer, and the obstacle of radius $a$ is seen as a smaller obstacle of radius $\delta$. In the case of a singular transformation ($\delta\approx0$), the inner boundary of $\omega$ would be mapped to a single point in $\Omega$, thereby eliminating any scattered waves.

\begin{figure*}
    \centering
    \subfloat[]{\includegraphics[height=0.27\textwidth,trim=50 5 50 0,clip]{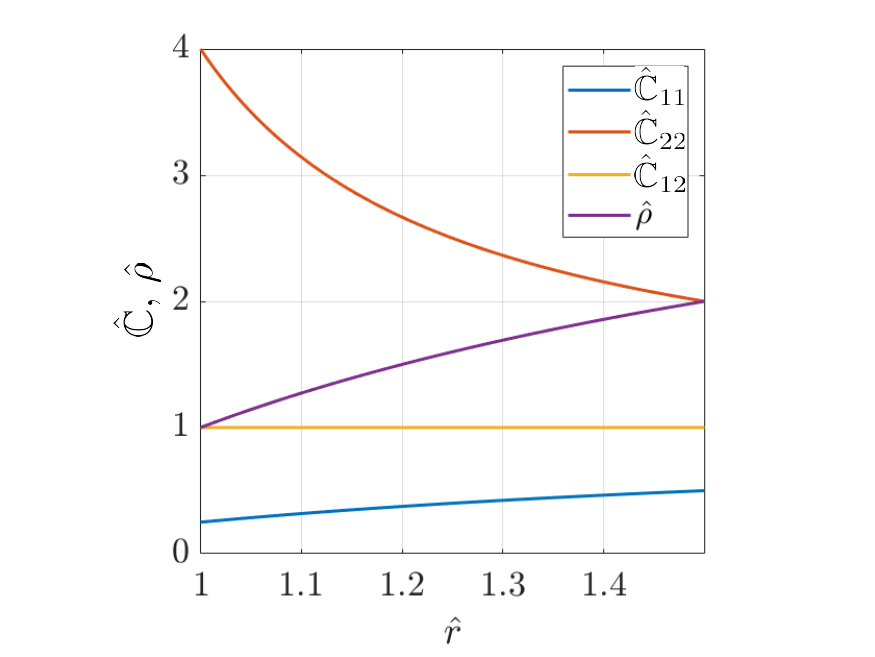}\label{fig:cloak_settings-a}}\qquad
    \subfloat[]{\includegraphics[height=0.25\textwidth]{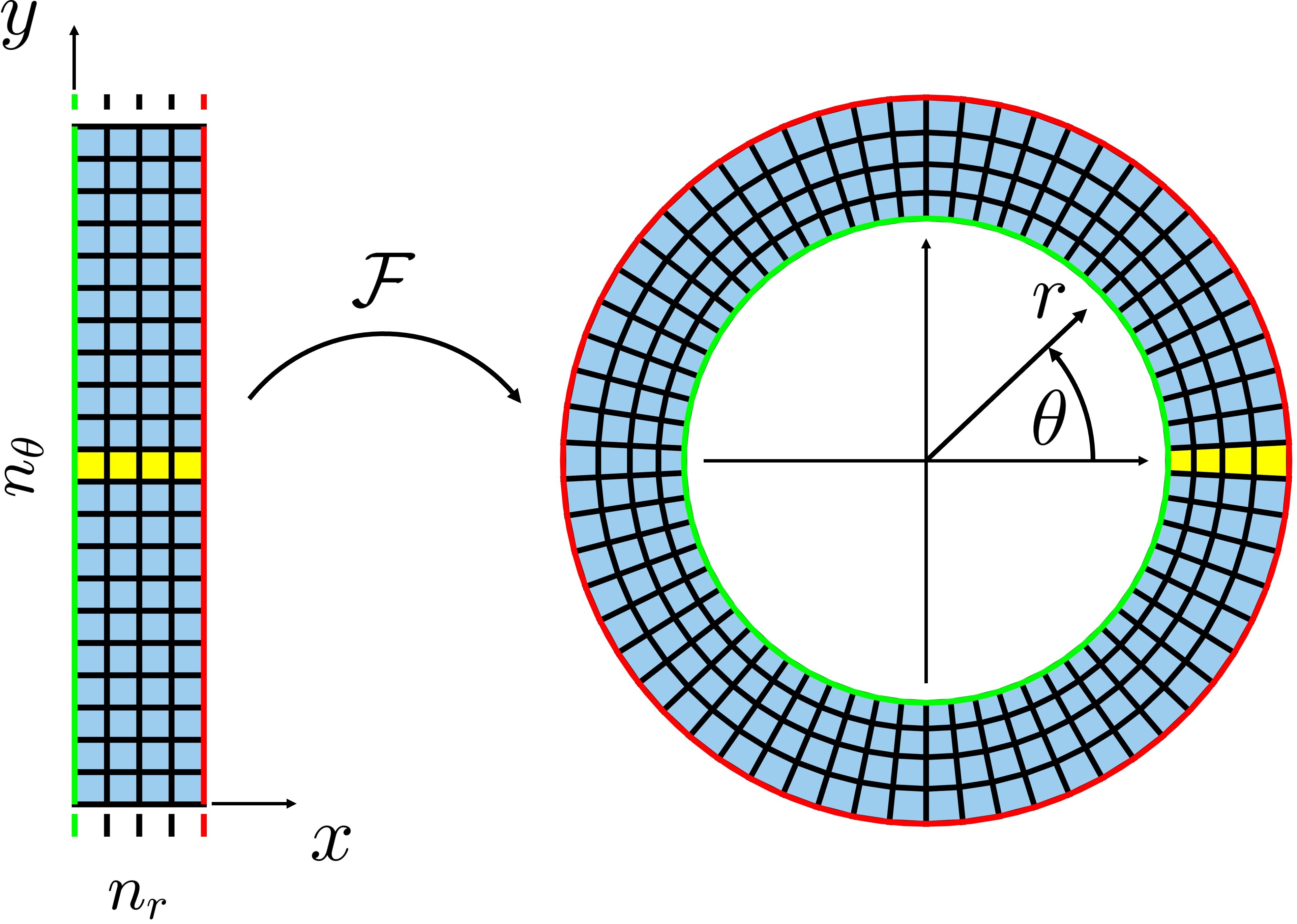}\label{fig:cloak_settings-b}}\qquad
    \subfloat[]{\includegraphics[height=0.25\textwidth]{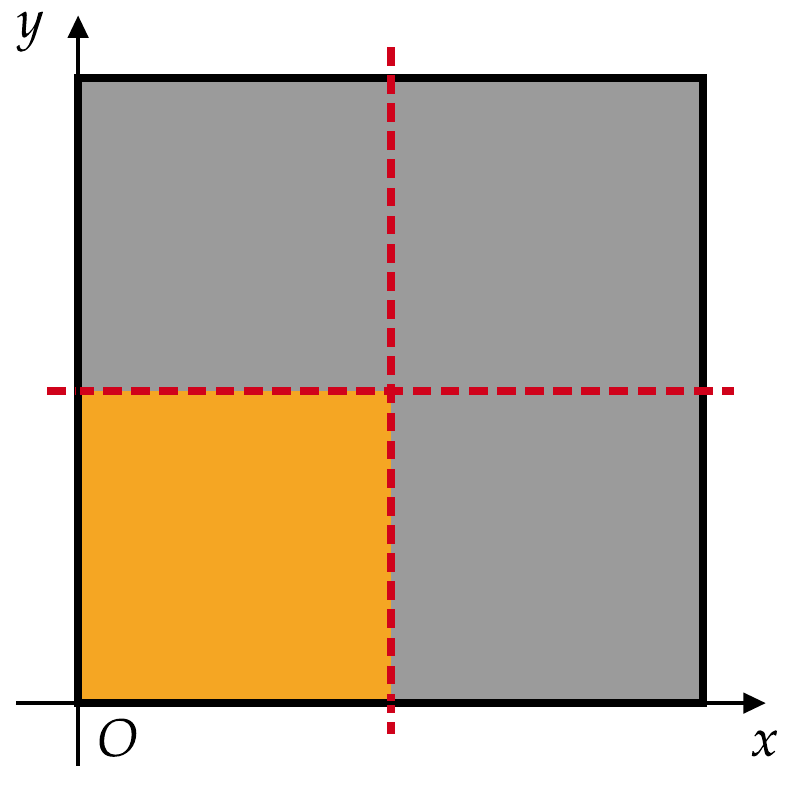}\label{fig:cloak_settings-c}}
    \caption{(a) Cloak target properties normalized with respect to water; (b) cloak discretization and microstructure map; and (c) cell symmetries with the primitive design domain in orange.
    }
    \label{fig:cloak_settings}
\end{figure*}

The cloak is made by a heterogeneous and anisotropic PM disposed around the object to conceal (Fig.~\ref{fig:cloak_settings-b}). Its elastic properties, normalized with respect to $\rho_0$ and $\kappa_0$, are shown in Fig.~\ref{fig:cloak_settings-a} as a function of the radius $\hat r$.
The elasticity tensor $\C$ given by Eq.~\eqref{eq:wave eq spatial} has the principal axis of anisotropy aligned with the radial and tangential directions of the circle. In particular, direction 1 is aligned with the radius.
The PM we optimize is made of square cells that fill the space according to a grid aligned with the orthogonal axis $x\times y$ (Fig.~\ref{fig:cloak_settings-b}).
We discretize the cloak using a polar grid, as shown in Fig.~\ref{fig:cloak_settings-b}, and project the Cartesian PM around the obstacle such that the $x$ axis aligns with the radial direction and the $y$ axis with the tangential. We use the following conformal map $\mathcal{F}$ to maintain the proportions of the cells
\begin{equation}\label{eq:conformal_map}
    \mathcal{F} \colon \begin{dcases}
        r = a\,\exp\Big(\frac{2\pi}{n_\theta\,l}\,x\Big)
        \\
        \theta =  \frac{2\pi}{n_\theta\,l}\,y
    \end{dcases}
\end{equation}
where $(x,y)\in[0,n_r\,l]\times[0,n_\theta\,l]$ is the space of the undeformed cartesian PM, $l$ is the cell edge, and $n_r=4$ and $n_\theta=62$ are the numbers of repetitions along the radial and tangential direction, respectively. The obstacle to conceal has radius $a=\SI{1}{\meter}$ and the virtual obstacle is chosen of radius $\delta=0.5a$. The external radius $b$ is determined by the conformal map \eqref{eq:conformal_map}
\begin{equation}
    b = a\,\exp\Big(\frac{2\pi \,n_\theta}{n_r}\Big) \approx 1.5\, a.
\end{equation}
In this way, we have several benefits: (i)~the vertical and horizontal axes of each cell align with the polar directions, (ii)~the contour of the cloak is well discretized, and (iii)~the symmetries of the cloak can be leveraged to have only 4 elementary cells.
Despite this deformation, the lattice can still be considered locally periodic in the approximation of cells much smaller than the radius of curvature. Such a procedure is quite standard and commonly accepted in the literature, see e.g.\ \cite{chen2017broadband,Quadrelli2021}.

\begin{figure*}
    \centering
    \includegraphics[width=0.8\textwidth]{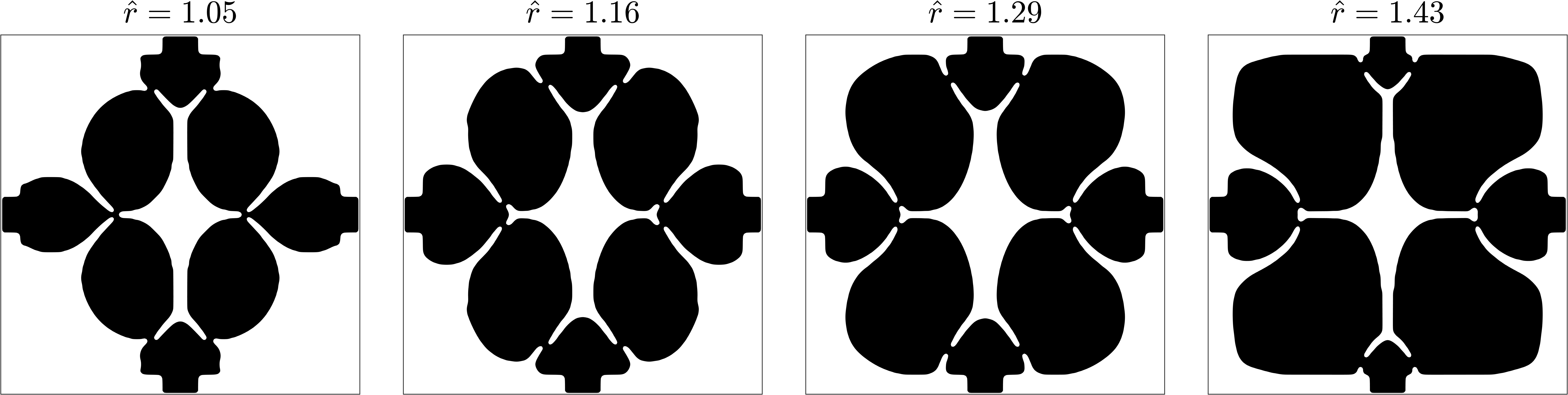}
    \caption{Optimized cloak cells obtained solving problem \eqref{opt:optimization_cloak}.}
    \label{fig:cloak_cells}
\end{figure*}

We then define the following optimization problem to design the anisotropic unit cells:
\begin{equation} \label{opt:optimization_cloak}
    \begin{aligned}
        \min_{\mu} \quad & C_{\mathrm{T}} & \\
        \mathrm{s.t.} \quad & 0.99 \, \C_{11}(\hat r) \leq \C^\mathrm{h}_{11} \leq 1.01 \, \C_{11}(\hat r) & \\
        & 0.99 \, \C_{22}(\hat r) \leq \C^\mathrm{h}_{22} \leq 1.01 \, \C_{22}(\hat r) & \\
        & 0.99 \, \C_{12}(\hat r) \leq \C^\mathrm{h}_{12} \leq 1.01 \, \C_{12}(\hat r) & \\
        & \C^\mathrm{h}_{33} \leq 0.01 \, \C_{12}(\hat r) & \\
        & 0.99 \, V_\mathrm{C}(\hat r) \leq V \leq 1.01 \, V_\mathrm{C}(\hat r) & \\
        & 0 \leq \mu \leq 1,
    \end{aligned}
\end{equation}
where the thermal compliance is used to avoid connectivity issues, and the double-sided constraints are treated as two separate constraints. In this section, the non-dimensional radius is defined as $\hat r\coloneqq r/a$.
\\
Using parallel computing with distributed memory over four processors via MPI \cite{walker1996mpi}, each cell was optimized in 400 iterations--enough to reach convergence--in approximately 5 minutes (0.75 seconds per iteration). The overall computational time for the four cells is around 20 minutes. The resulting density distributions are shown in Fig.~\ref{fig:cloak_cells}.

\begin{figure*}
    \centering
    \subfloat[]{\includegraphics[height=0.3\textwidth,trim=25 0 25 0,clip]{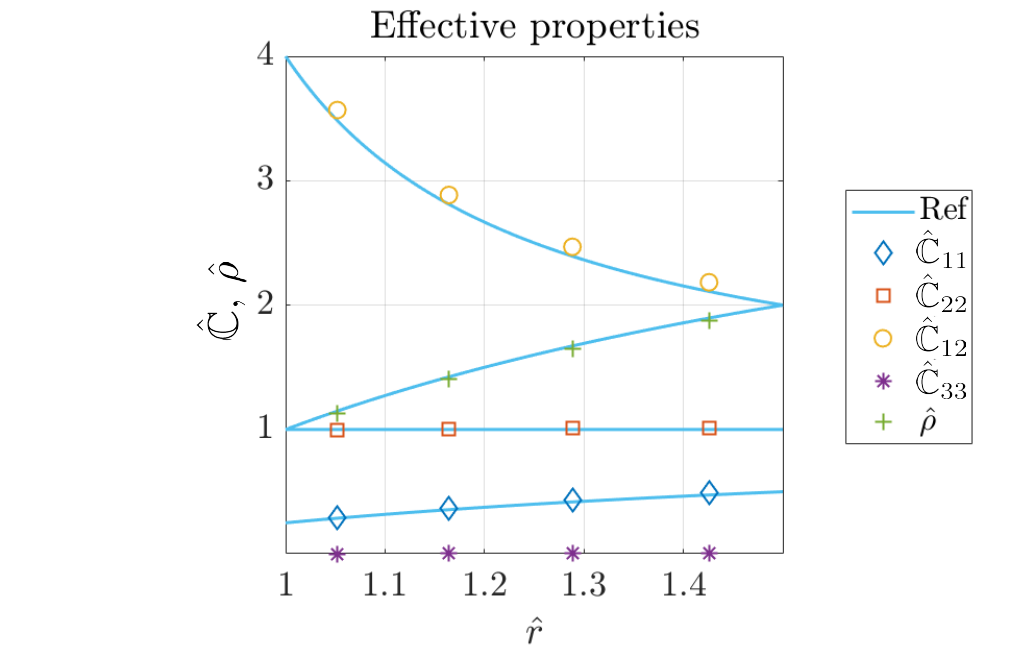}}
    \subfloat[]{\includegraphics[height=0.3\textwidth,trim=25 0 25 0,clip]{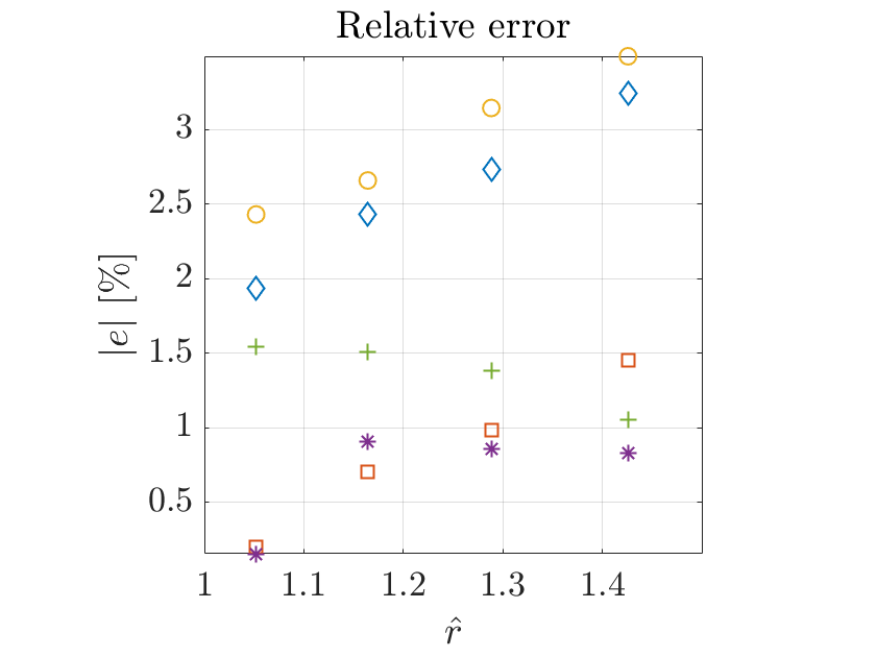}}
    \caption{(a) the effective properties (normalized with respect to water) of the optimized cloak cells in Fig.~\ref{fig:cloak_cells}, computed using \Comsol{}; (b) the magnitude of the relative error between effective and target properties.
    }
    \label{fig:cloak_properties}
\end{figure*}

\subsection{Dynamic analysis} \label{sec:dyn_analysis}

\begin{figure*}
    \centering
    \subfloat[]{\includegraphics[height=0.285\textwidth,trim=30 13 100 0,clip]{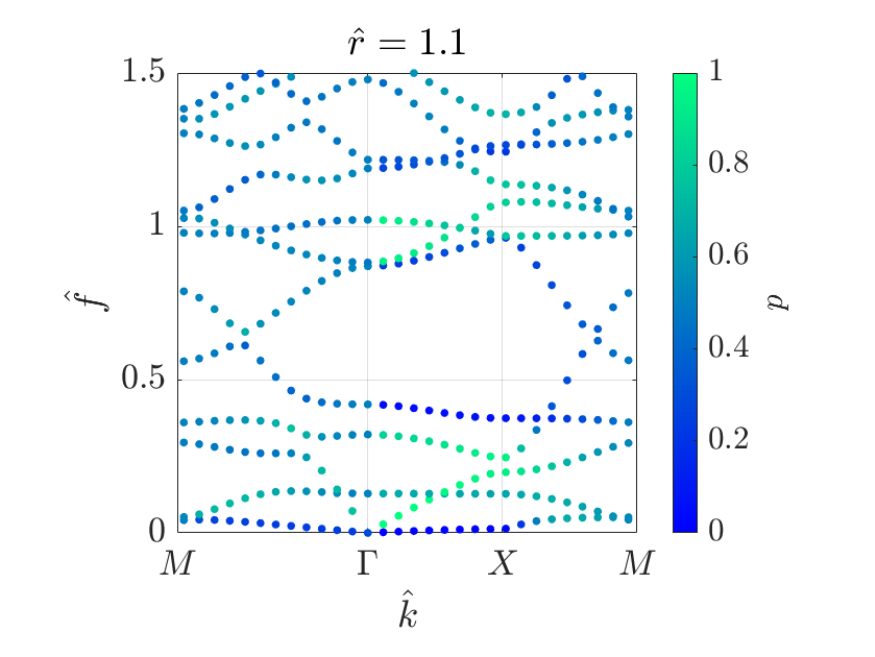}}
    \subfloat[]{\includegraphics[height=0.285\textwidth,trim=63 13 100 0,clip]{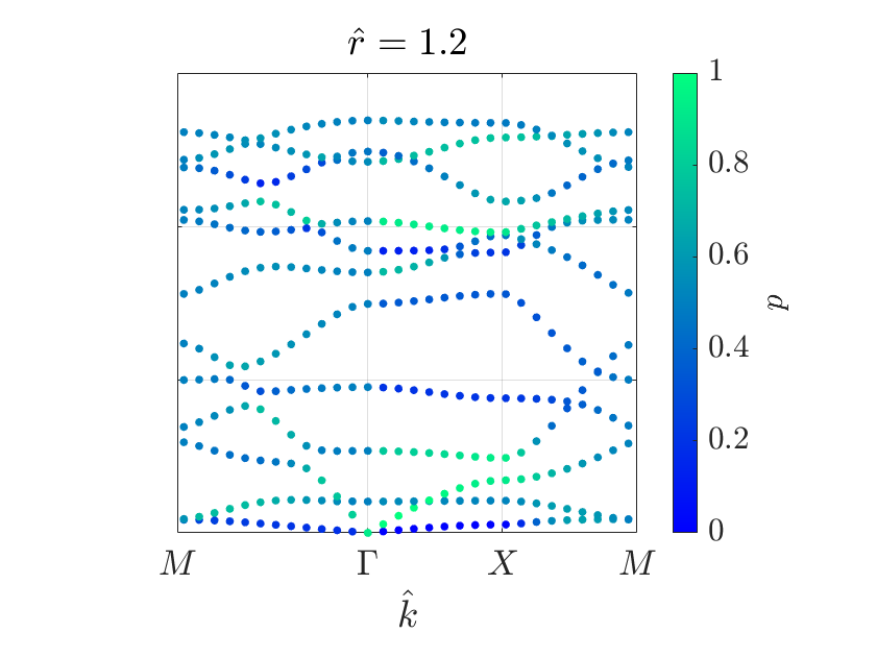}}
    \subfloat[]{\includegraphics[height=0.285\textwidth,trim=63 13 100 0,clip]{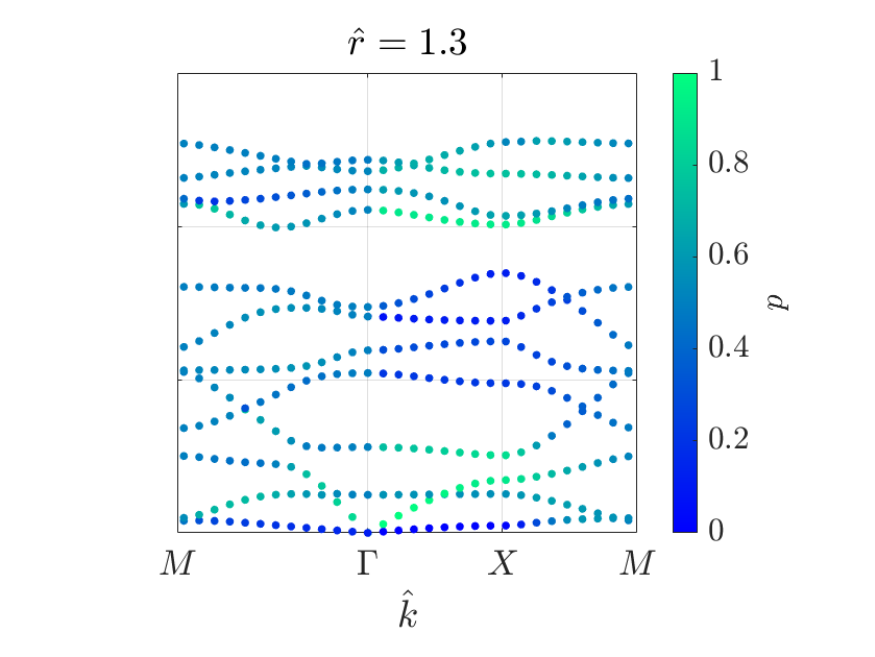}}
    \subfloat[]{\includegraphics[height=0.285\textwidth,trim=63 13 60 0,clip]{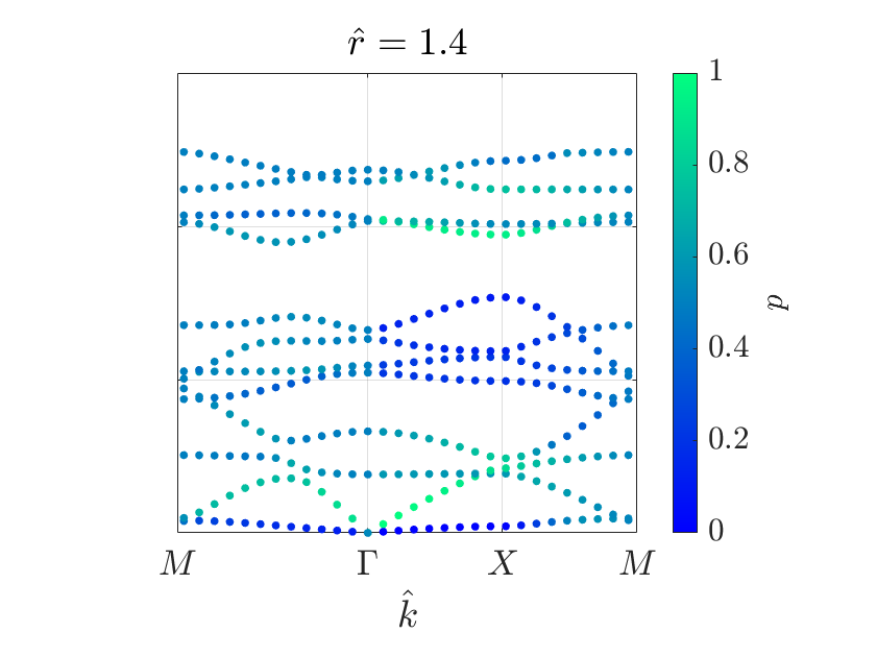}}
    \caption{Dispersion diagrams of the optimized cloak cells in Fig.~\ref{fig:cloak_cells}, computed using \Comsol{}.
    }
    \label{fig:cloak_dispersion}
\end{figure*}

The optimized cloak cells are imported into \Comsol{} to evaluate their effective properties using a body-fitted mesh. The comparison provided in Fig.~\ref{fig:cloak_properties} shows that the magnitude of the relative error between simulated and target properties does not exceed $3.5\%$, which is considered acceptable for the proposed application.
\\
To assess the dynamic properties, we also compute the dispersion diagrams of the four cells shown in Fig.~\ref{fig:cloak_dispersion} with the polarization color scale defined by Eq.~\eqref{eq:polariz}. In each diagram, two acoustic branches depart from the point $\Gamma$: one has a slope much smaller than the other because the former is related to a transversal shear wave, the latter to a longitudinal pressure wave. Although there is no pentamode bandgap, the following simulations show that the energy is mainly carried by pressure waves.

\begin{figure*}
    \centering
    \begin{subfigure}{.32\textwidth}
        \centering
        \includegraphics[width=\textwidth,trim=55 10 45 20,clip]{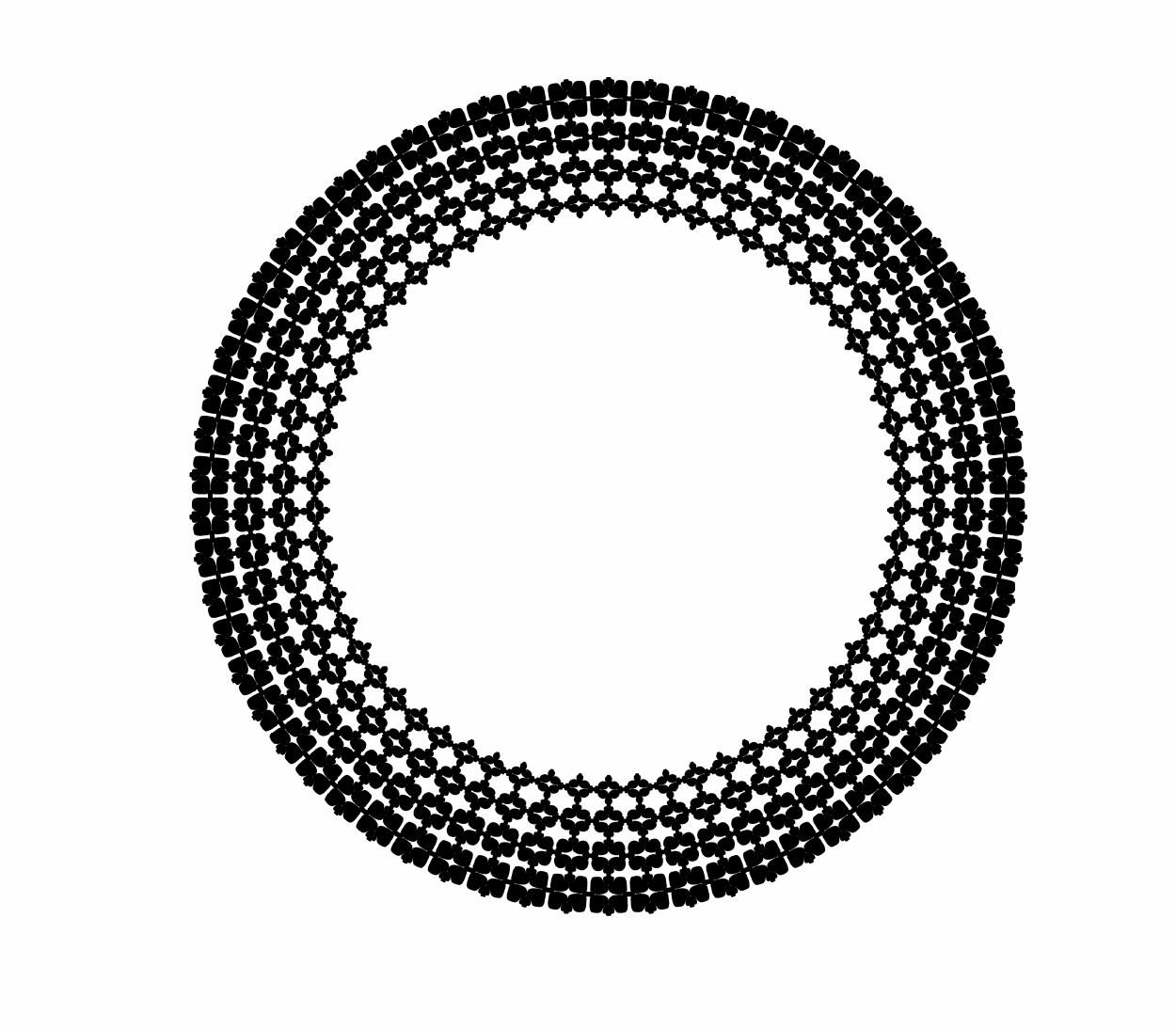}
        \caption{}\label{fig:cloak_pressure-a}
    \end{subfigure}
    \begin{subfigure}{.33\textwidth}
        \centering
        \begin{tikzpicture}
            \matrix[matrix of nodes,
                    row sep=-1pt, column sep=-1pt,
                    nodes={inner sep=0, anchor=center}] (main) {
            \includegraphics[width=0.49\textwidth,trim=70 25 70 0,clip]{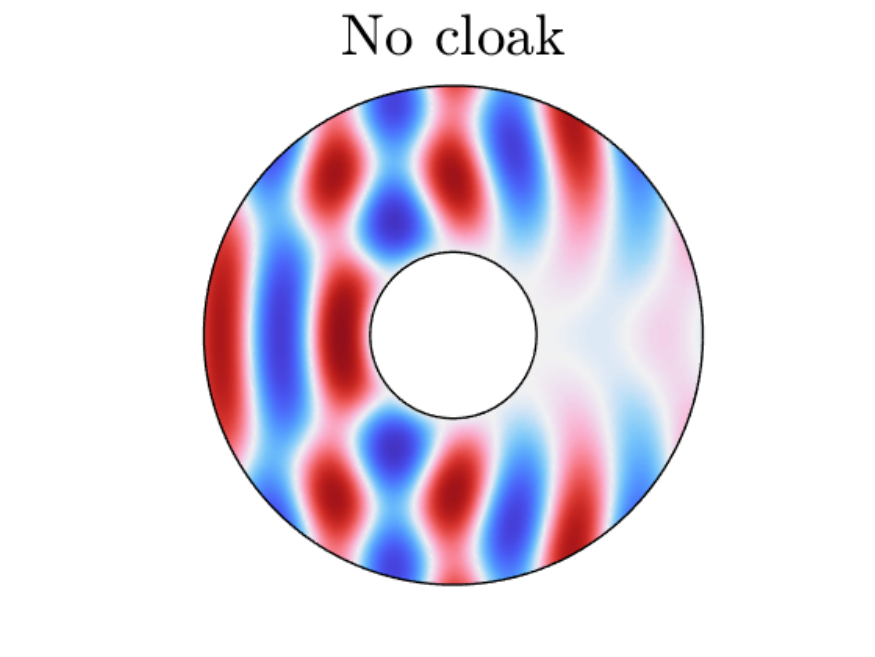} &
            \includegraphics[width=0.49\textwidth,trim=70 25 70 0,clip]{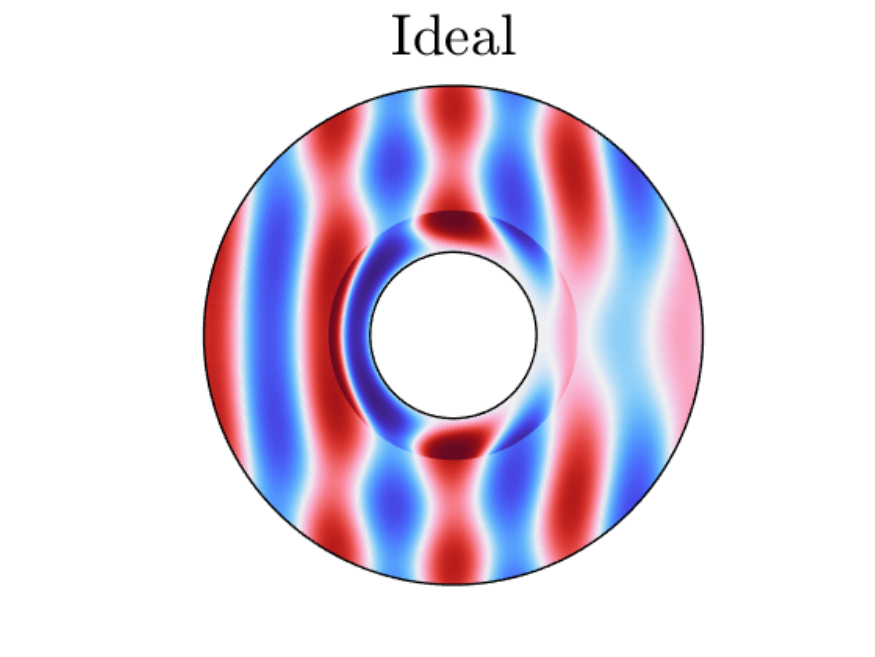} \\
            \\
            \includegraphics[width=0.49\textwidth,trim=70 25 70 0,clip]{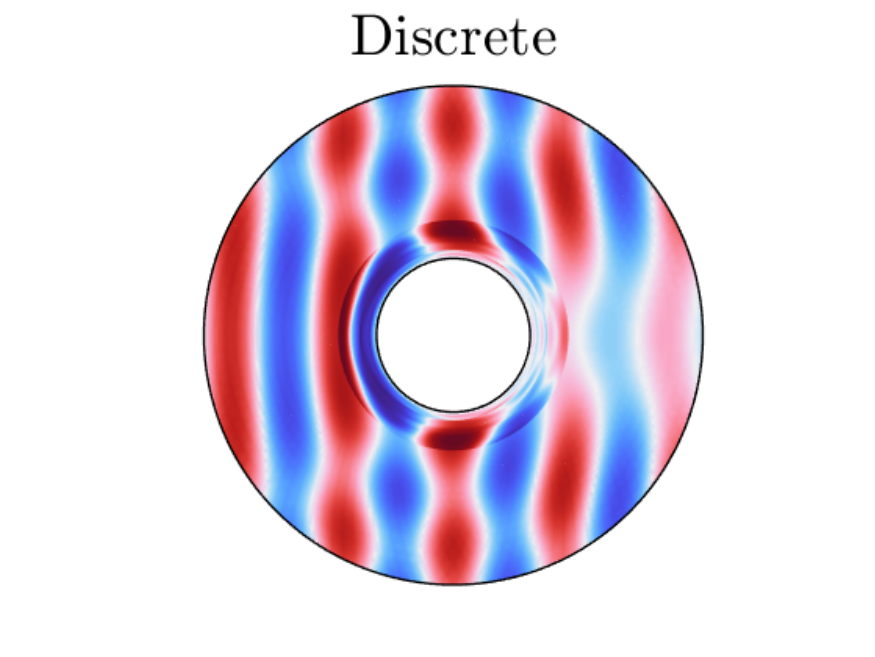} & 
            \includegraphics[width=0.49\textwidth,trim=70 25 70 0,clip]{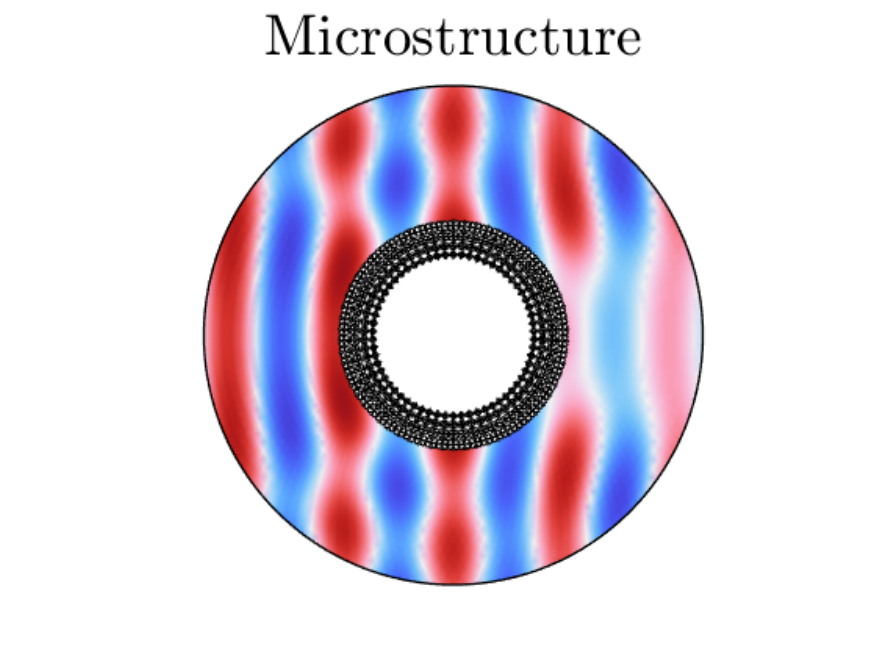} \\
            };
            \node[anchor=center] at (main.center)
                {\includegraphics[width=0.12\textwidth,trim=-20 0 0 -20]{Colorbar_pressure.png}};
        \end{tikzpicture}
        \caption{}\label{fig:cloak_pressure-b}
    \end{subfigure}
    \begin{subfigure}{.3\textwidth}
        \centering
        \includegraphics[width=\textwidth,trim=60 0 60 0,clip]{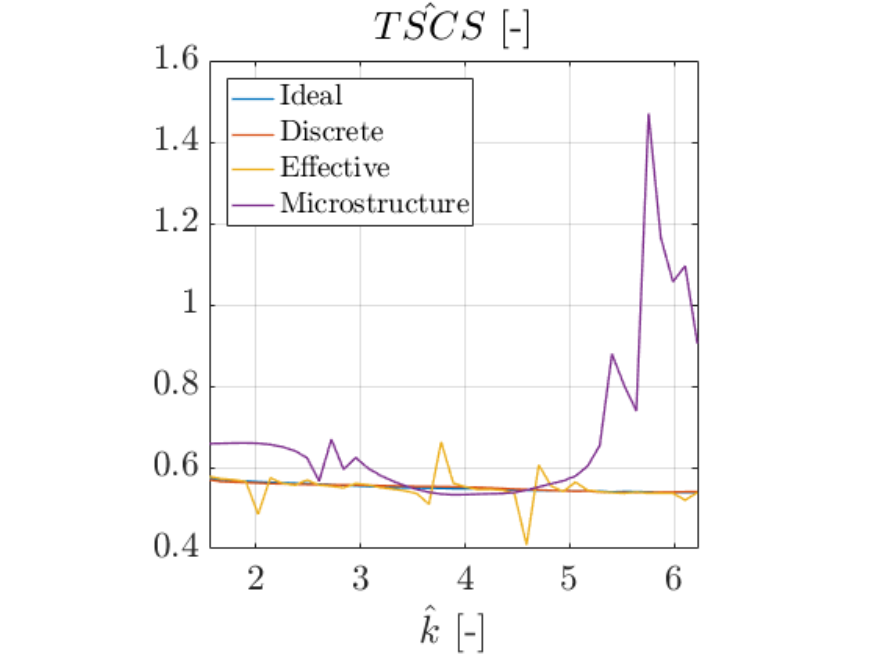}
        \caption{}\label{fig:cloak_pressure-c}
    \end{subfigure}
    \caption{(a) Cloak microstructure, (b) real part of the pressure field phasor for a plane wave incident on a soft obstacle without cloak compared with subsequent levels of discretization: the ideal cloak, the discrete ideal cloak and the microstructured cloak. (c) Normalized Total Scattering Cross Section ($\hat{\mathrm{TSCS}}$) of the four scenarios over the non dimentional wavenumber $\hat k\coloneqq k\, a\in[\pi/2, 2\pi]$.
    }
    \label{fig:cloak_pressure}
\end{figure*}

The whole microstructure is assembled as shown in Fig.~\ref{fig:cloak_pressure-a} and simulations are performed in the frequency domain to assess the performance of the cloak. In Fig.~\ref{fig:cloak_pressure-b}, a soft circular obstacle (zero-pressure boundaries) is probed by a plane wave that comes from left with a non-dimensional frequency of $\hat f = f\; l/c_0 = 0.07$
(here, the edge of the inner cells $l=a/n_\theta$ is considered).
This corresponds to a wavelength $\lambda = 1.4\, a$, 
and an important scattered field is expected.
Four scenarios are considered: the circular obstacle without a cloak, the obstacle equipped by an ideal cloak, the obstacle with a discrete ideal cloak and the obstacle with the microstructured cloak.

The scattering reduction is clearly visible in all three cloaked configurations. A quantitative comparison is obtained by evaluating the so-called Total Scattering Cross Section (TSCS) \cite{quadrelli2025reduced}. Figure~\ref{fig:cloak_pressure-c} reports the non-dimensional TSCS of the cloaked obstacle, normalized with respect to that of the uncloaked obstacle,
\begin{equation}
    \hat{\mathrm{TSCS}}_\mathrm{clk} = \frac{\mathrm{TSCS}_\mathrm{clk}}{\mathrm{TSCS}_\mathrm{obs}},
\end{equation}
so that values below unity indicate an effective cloaking performance.
The frequency range considered is such that the non-dimensional wavenumber is $\hat k\coloneqq k\, a\in[0.5\pi, 2\pi]$.
The ideal and discrete configurations exhibit almost overlapping results within this frequency range. The cloak based on the effective properties of the microstructure performs similarly, although narrow spikes appear, likely due to the small shear moduli of the unit cells. The fully microstructured cloak yields a comparable $\hat{\mathrm{TSCS}}_\mathrm{clk}$, generally slightly higher than the others. Around $\hat{k}=5.7$, however, its performance becomes worse than that of the bare obstacle. This behavior is attributed to the breakdown of the homogenization assumption, as the wavelength becomes comparable to the characteristic size of the microstructure. Improved performance at higher frequencies could be achieved with a finer discretization, but this lies beyond the scope of the present study and is a well-known effect in the literature.

\section{Conclusion} \label{sec:conclusion}

We presented a robust and efficient topology optimization framework for designing pentamode-based acoustic metamaterials. Unlike traditional parametric optimization, our approach does not require a predefined base geometry. This significantly increases design flexibility and avoids potential geometric conflicts. We incorporated the Virtual Temperature Method (VTM) to ensure that the designs are physically consistent and manufacturable, thereby mitigating connectivity issues that commonly arise in low-stiffness layouts.

The optimized unit cells and full microstructures were validated by solving the coupled acoustic-elastic problem using \Comsol{}, which confirmed their acoustic performance. These results demonstrate the effectiveness of topology optimization in the automated design of complex acoustic metamaterials. This method offers a scalable, flexible alternative to conventional parametric approaches, ensuring connectivity and full-wave functionality.

Future research will extend the methodology to three-dimensional geometries and address challenges such as undercuts and full anisotropy. These developments will further expand the scope of the framework, enabling the design of more complex and versatile metamaterial structures for advanced engineering applications.

%
%
%
%
%
%


\section*{Replication of results}

All the topology optimization routines, including the optimized results and the \Comsol{} files used for the analysis, are available on GitHub: \url{https://github.com/matteo12pozzi/luneburg-lens}.

\appendix

\section{Sensitivity analysis} \label{app:sensitivity}

Using the chain rule, the derivative of the effective volume $V$ with respect to the filtered density $\tilde\mu$ is
\begin{equation} \label{eq:pV_muFilt}
    \mathcal{D}_{\tilde\mu} V [\tilde\psi] = \int_{\Omega} \mathcal{D}_{\tilde\mu} \bar{\mu} [\tilde\psi] \, \mathrm{d}x,
\end{equation}
where $\mathcal{D}_{\tilde\mu} V [\tilde\psi]$ indicates the \textit{Gateaux derivative} \cite{Manzoni2021optimal} of $V$ with respect $\tilde\mu$ in the direction of the test function $\tilde\psi \in \tilde{\mathcal{M}}$. Likewise, the term $\mathcal{D}_{\tilde\mu} \bar{\mu} [\tilde\psi]$ is the derivative of the projected density field $\bar\mu$ with respect to $\tilde\mu$ in the direction $\tilde\psi$:
\begin{equation} \label{eq:pmuProj_muFilt}
    \mathcal{D}_{\tilde\mu} \bar{\mu} [\tilde\psi] = \frac{1 - \tanh^2{(\beta (\tilde\mu - \eta))}}{\tanh(\beta \eta) + \tanh(\beta(1 - \eta))} \beta \tilde\psi.
\end{equation}

The adjoint method \cite{Manzoni2021optimal} is used to obtain the derivative of $\C^\mathrm{h}_{ij}$ with respect to $\tilde{\mu}$ as in \cite{Pozzi2024luneburg}:
\begin{equation} \label{eq:pCh_muFilt}
    \mathcal{D}_{\tilde\mu} \C^\mathrm{h}_{ij}[\tilde\psi] = \frac{1}{V_0} \int_\Omega{ \mathcal{D}_{\tilde\mu} (\hat\mu \circ \bar\mu) [\tilde\psi] \,(\bs{\varepsilon}_i^0 - \bs{\varepsilon}_i) : \C :(\bs{\varepsilon}_j^0 - \bs{\varepsilon}_j) \,\mathrm dx},
\end{equation}
where $\mathcal{D}_{\tilde\mu} (\hat\mu \circ \bar\mu) [\tilde\psi]$ is obtained by differentiating Eqs.~\eqref{eq:projection_scheme} and~\eqref{eq:simp_scheme} using the chain rule:
\begin{equation} \label{eq:pmuSimp_muFilt}
\begin{aligned}
    \mathcal{D}_{\tilde\mu} (\hat\mu \circ \bar\mu) [\tilde\psi] &= \mathcal{D}_{\bar\mu} \hat\mu [\mathcal{D}_{\tilde\mu} \bar\mu [\tilde\psi]]
    \\
    &= p (1 - \hat\mu_0)\, \bar\mu^{p - 1} \frac{1 - \tanh^2{(\beta (\tilde\mu - \eta))}}{\tanh(\beta \eta) + \tanh(\beta(1 - \eta))} \beta \tilde\psi.
\end{aligned}
\end{equation}

The sensitivity of $\C_{ij}^\mathrm{h}$ with respect to the density $\mu$ is computed using the adjoint method \cite{Manzoni2021optimal}. In particular, the Lagrangian function is defined as
\begin{equation} \label{eq:lagrangian_Ch}
    \mathcal{L}(\mu, \tilde\mu, \lambda) = \C_{ij}^\mathrm{h} + \mathcal{D}_{\tilde\psi} h [\lambda].
\end{equation}
The adjoint variable $\lambda \in \tilde{\mathcal{M}}$ is computed from the adjoint equation below:
\begin{equation} \label{eq:adjoint_equation_Ch}
    \mathcal{D}_{\tilde\mu} \mathcal{L} [\tilde\psi] = \mathcal{D}_{\tilde\mu} \C_{ij}^\mathrm{h} [\tilde\psi] + \int_{\Omega} R_f^2 \, \nabla \tilde\psi \cdot \nabla \lambda \, \mathrm{d}x +\int_{\Omega} \tilde\psi \, \lambda \, \mathrm{d}x = 0,
\end{equation}
where $\mathcal{D}_{\tilde\mu} \C_{ij}^\mathrm{h} [\tilde\psi]$ is computed from Eq.~\eqref{eq:pCh_muFilt}. After solving for $\lambda$, the sensitivity of $\C_{ij}^\mathrm{h}$ is the partial derivative of $\mathcal{L}$ with respect to $\mu$:
\begin{equation} \label{eq:sensitivity_Ch}
    \mathcal{D}_{\mu} \mathcal{L} [\psi] = -\int_{\Omega} \psi \, \lambda \, \mathrm{d}x,
\end{equation}
where $\psi \in \mathcal{M}$ is the test function.
\\
The same approach is used to compute the sensitivity of $V$ with respect to $\mu$:
\begin{align} \label{eq:sensitivity_V}
    \mathcal{D}_{\tilde\mu} \mathcal{L} [\tilde\psi] &= \mathcal{D}_{\tilde\mu} V [\tilde\psi] + \int_{\Omega} R_f^2 \, \nabla \tilde\psi \cdot \nabla \lambda \, \mathrm{d}x +\int_{\Omega} \tilde\psi \, \lambda \, \mathrm{d}x = 0,
    \\
    \mathcal{D}_{\mu} \mathcal{L} [\psi] &= -\int_{\Omega} \psi \, \lambda \, \mathrm{d}x.
\end{align}
In the same way, the sensitivity of the thermal compliance $C_{\mathrm{T}}$ with respect to the density $\mu$ is computed from
\begin{align} \label{eq:sensitivity_Ct}
    \mathcal{D}_{\tilde\mu} \mathcal{L} [\tilde\psi] &= \mathcal{D}_{\tilde\mu} C_{\mathrm{T}} [\tilde\psi] + \int_{\Omega} R_f^2 \, \nabla \tilde\psi \cdot \nabla \lambda \, \mathrm{d}x +\int_{\Omega} \tilde\psi \, \lambda \, \mathrm{d}x = 0,
    \\
    \mathcal{D}_{\mu} \mathcal{L} [\psi] &= -\int_{\Omega} \psi \, \lambda \, \mathrm{d}x,
\end{align}
where the derivative of $C_{\mathrm{T}}$ with respect to $\tilde \mu$ is computed using the adjoint method as in \cite{Pozzi2024luneburg}:
\begin{equation} \label{eq:pCt_muFilt}
    \mathcal{D}_{\tilde\mu} C_{\mathrm{T}} [\tilde\psi] = -\int_{\Omega} \mathcal{D}_{\tilde\mu} (\hat\mu \circ \bar\mu) [\tilde\psi] \kappa_{\mathrm{T}} \nabla T \cdot \nabla T \, \mathrm{d}x.
\end{equation}
The sensitivities are efficiently implemented in their weak form using FEniCSx \cite{fenicsx}.

\bibliographystyle{elsarticle-num-names}
\bibliography{bibliography}

\end{document}